\documentclass[fleqn,9pt]{SelfArx} 

\definecolor{color1}{RGB}{0,0,90} 
\definecolor{color2}{RGB}{0,20,20} 

\usepackage{hyperref} 
\usepackage{epstopdf}
\usepackage[alsoload=synchem, detect-weight=true, detect-family=true]{siunitx}[=v2]
\usepackage[superscript]{cite}
\usepackage[figurename=Fig.,justification=justified]{caption}
\usepackage{plain}
\usepackage[normalem]{ulem}
\usepackage{upgreek}
\hypersetup{hidelinks,colorlinks,breaklinks=true,urlcolor=color2,citecolor=color1,linkcolor=color1,bookmarksopen=false,pdftitle={Title},pdfauthor={Author}}
\DeclareSIUnit{\molar}{M}

\JournalInfo{23 December 2025} 
\Archive{} 

\PaperTitle{Funnelling super-resolution STED microscopy through multimode fibres} 
\newcommand*\samethanks[1][\value{footnote}]{\footnotemark[#1]}
\Authors{Andr\'{e} D. Gomes\textsuperscript{1}\thanks{correspondence to andre.gomes@leibniz-ipht.de \& tomas.cizmar@leibniz-ipht.de},
Miroslav Stib\r{u}rek\textsuperscript{2},
Sergey Turtaev\textsuperscript{1},
Tom\'{a}\v{s}  Pik\'{a}lek\textsuperscript{2},
Katharina Reglinski\textsuperscript{1,3,4},
Christian Eggeling\textsuperscript{1,3,5,6},
\&~Tom\'{a}\v{s} \v{C}i\v{z}m\'{a}r\textsuperscript{1,2,3}\samethanks}

\affiliation{\textsuperscript{1}\textit{Leibniz Institute of Photonic Technology, Albert-Einstein-Stra{\ss}e 9, 07745 Jena, Germany}} 
\affiliation{\textsuperscript{2}\textit{Institute of Scientific Instruments of the Czech Academy of Sciences, Kr\'{a}lovopolsk\'{a} 147, 612 64 Brno, Czech Republic}}
\affiliation{\textsuperscript{3}\textit{Institute of Applied Optics and Biophysics, Friedrich Schiller University Jena, Max-Wien-Platz 1, 07743 Jena, Germany}} 
\affiliation{\textsuperscript{4}\textit{University Hospital Jena, Bachstra{\ss}e 18, 07743 Jena, Germany}} 
\affiliation{\textsuperscript{5}\textit{Human Immunology Unit, Weatherall Institute of Molecular Medicine, University of Oxford, Headley Way, Oxford, OX3 9DS, United Kingdom}} 
\affiliation{\textsuperscript{6}\textit{Jena Center for Soft Matter, Philosophenweg 7, 07743 Jena, Germany}} 

\newcommand{\keywordname}{Keywords} 

\usepackage{xcolor}

\definecolor{cincinnati-green}{RGB}{0,190,0}

\Abstract{\bf 
Holographic multimode fibre endoscopes have recently shown their ability to unveil and monitor deep brain structures with sub-micrometre resolution, establishing themselves as a minimally-invasive technology with promising applications in neurobiology. In this approach, holographic control of the input light field entering the multimode fibres is achieved by means of wavefront shaping, usually treating the fibre as a complex medium. 
In contrast to other unpredictable and highly scattering complex media, multimode fibres feature symmetries and strong correlations between their input and output fields. 
Both step-index and graded-index multimode fibres offer a specific set of such correlations which, when appropriately leveraged, enable generating high-quality focused pulses with minimal intermodal dispersion. 
With this, we funnelled pulsed super-resolution STED microscopy with time-gated detection through a custom multimode fibre probe, combining the correlations of both multimode fibre types. We demonstrate resolution improvements over 3-times beyond the diffraction limit and showcase its applicability in bioimaging. This work provides not only a solution for delivering short pulses through step-index multimode fibre segments but also marks a step towards bringing advanced super-resolution imaging techniques with virtually no depth limitations.
}

\begin{document}
\thispagestyle{empty} 
\sloppy
\fontsize{3.3mm}{4.3mm}\selectfont
\flushbottom 
\maketitle 

\addcontentsline{toc}{section}{Introduction} 

\noindent
Gaining insight into the mechanisms underlying brain physiology is essential for advancing our understanding of brain diseases. This quest demands technologies capable of enabling high resolution imaging, particularly at greater depths and with minimal damage \cite{Uhlirova2024}, in order to study neuronal circuits in detail. Specifically, observing complex structures such as dendritic spines and their plasticity in in-vivo brain tissue provides insights on cell-to-cell communication. As an example, small abnormalities in spine morphology can severely impact brain function, as in the case of brain disorders such as Alzheimer's disease and epilepsy \cite{Arizono2023}.

In this regard, the rapidly evolving field of holographic endoscopy offers a promising solution to observe deep brain structures and connectivity with sub-micrometre resolution \cite{Stiburek2023}. Here, multimode optical fibres (MMFs) are utilised as minimally-invasive endoscopes, relying on holographic wavefront shaping to gain precise control over the light transport. At the power levels used in imaging applications, the seemingly disordered nature of light propagation within MMFs remains almost lossless and linear, and therefore deterministic. This way the relationship between a set of light fields entering and leaving the MMF segment can be described and empirically measured in the form of a transmission matrix (TM), the digital twin of the MMF-comprising optical system \cite{Popoff2010, Popoff2011, Cizmar2011a, Kim2015}. Once available, the TM enables the synthesis of any desired optical landscape within the constraints of the MMF \cite{Popoff2011}. Such output light fields, particularly diffraction-limited foci, can be generated with high quality through appropriate modulation and control of two input orthogonal polarisation states \cite{Gomes2022}. The ability to holographically control and shape light through MMFs opened up a route to various biophotonics applications, such as holographic optical tweezers \cite{Leite2018a} and scanning imaging techniques, particularly fluorescence microscopy \cite{Cizmar2012, Papadopoulos2013a, Turtaev2018c, Vasquez-Lopez2018, Wen2023}. 
Breaking through the current state-of-the-art resolution of holographic endoscopes would create possibilities for detailed observation of small structures and their plasticity in in-vivo brain tissue, especially structural manifestation of synapses, dendritic spines, with dimensions as small as \SI{230}{\nano\metre}\cite{Li2023}.

In the previously described applications, MMFs were treated as entirely random media, however their light transport is not genuinely random. In fact, MMFs exhibit rich properties and symmetries that shape and influence how light propagates through them. A clear manifestation of this are the strong correlations between the input and output fields of the fibre \cite{Li2021}. By harnessing and carefully tailoring these input-output correlations, it is possible to address some of the significant challenges in the field. One such challenge is the ability to focus and deliver short pulses without intermodal dispersion, unlocking advanced imaging techniques that rely on pulsed laser sources, such as super-resolution Stimulated Emission Depletion (STED) fluorescence microscopy \cite{Hell1994}. Nowadays, super-resolution techniques and especially STED microscopy are routinely employed for tissue imaging, yet so far utilising objective lenses with adapted immersion media such as glycerol \cite{Urban2011, Berning2012} or adaptive optics \cite{Zdankowski2019, Velasco2021, Hao2021}. Pulsed STED microscopy is directly compatible with holographic MMF endoscopes, since it is, in its conventional form, relying on scanning tightly focussed spots over the sample to produce the final image. Nonetheless, the successful realisation of pulsed STED microscopy through an MMF requires fulfilling essential conditions, including the elimination or suppression of intermodal dispersion and the delivery of high-quality output beams, particularly for the depletion beam \cite{Blom2017}.

In this paper, we combine the conservation of the propagation constant in step-index MMFs \cite{Cizmar2012} with the self-imaging property of graded-index MMFs \cite{Agrawal1974, Iga1980}. The resulting `endcap fibre' lends itself to synthesise higher-quality focused pulses than would be possible by using either fibre type individually. This approach enabled us to funnel pulsed STED microscopy with time-gated detection through a holographic fibre endoscope. We demonstrate a resolution improvement over 3-fold beyond the diffraction limit and showcase the applicability of the concept in bio-imaging. This work is a step towards offering highly sought-after super-resolution imaging methods with virtually no depth limitation. 

\subsection*{Results}
\subsubsection*{Wavefront shaping of pulsed light through multimode fibres}

\begin{figure*}[ht!]
   \centering
   \includegraphics[width=\textwidth]{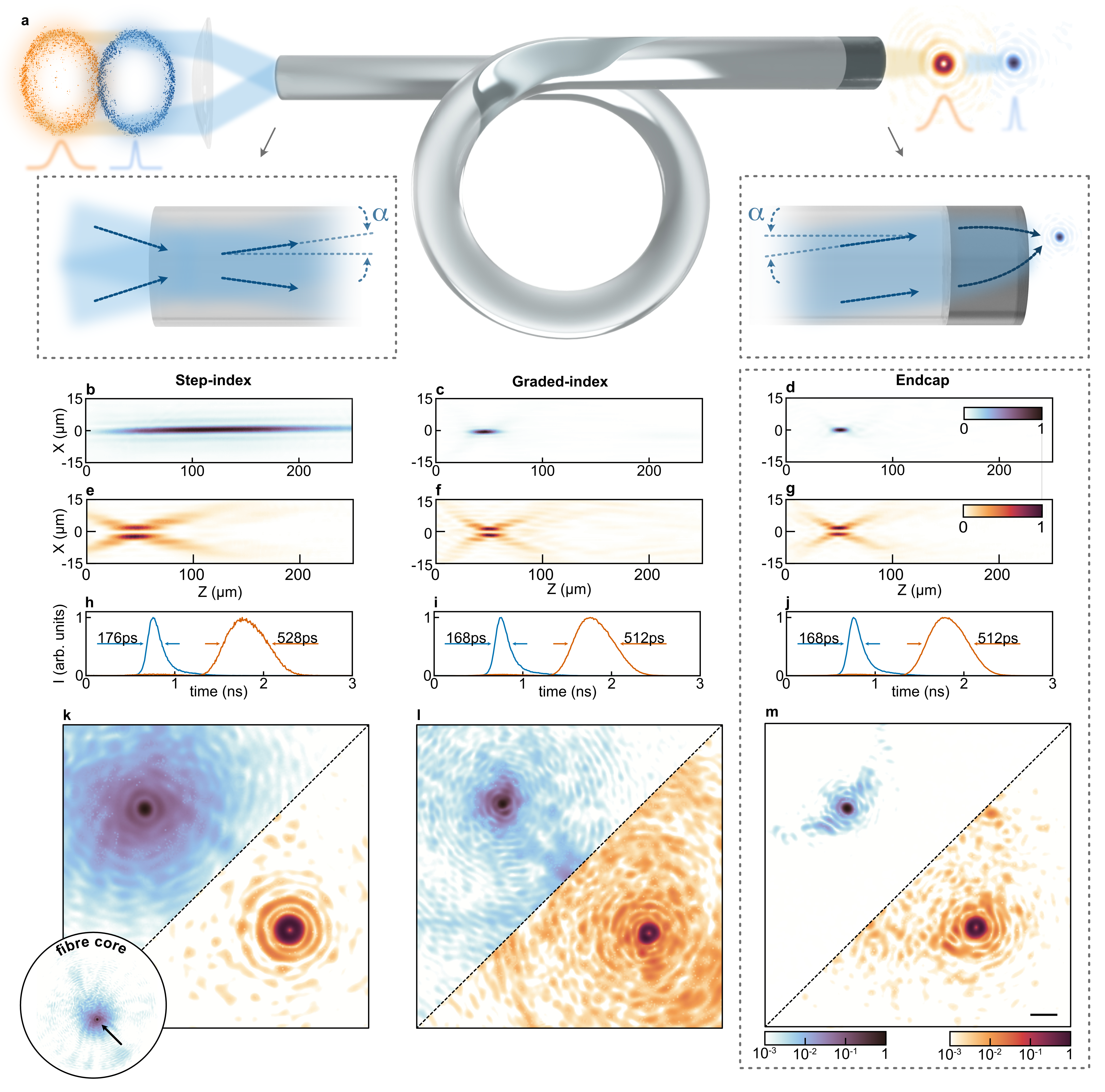} 
   \caption{{\bf Focussing and shaping pulsed laser light through an endcap multimode fibre (MMF) endoscope.} 
{\bf a}, Illustration of the endcap probe working principle. Light from an annular region couples and propagates through a step-index MMF segment mainly within a single mode group (with conserved propagation angle, $\alpha$), being focussed at the output by the graded-index MMF endcap. 
{\bf b-d}, Experimental axial extent of a \SI{485}{\nano\metre} pulsed laser focussed through a \SI{105}{\micro\metre}-diameter core and $0.22$ NA step-index multimode fibre, a \SI{100}{\micro\metre}-diameter core and $0.29$ NA graded-index multimode fibre, and a \SI{100}{\micro\metre}-diameter core and $0.29$ effective NA endcap fibre probe, respectively. 
{\bf e-g}, Experimental axial extent of a \SI{592}{\nano\metre} pulsed laser vortex beam focussed through the step-index, graded-index, and endcap fibre probes, respectively. 
{\bf h-j} Experimental temporal profile of a \SI{485}{\nano\metre} (blue) and a \SI{592}{\nano\metre} (orange) picosecond pulsed lasers focussed through the step-index, graded-index, and endcap fibre probes, respectively. 
{\bf k-m} High dynamic range images, in logarithmic scale, of the pulsed excitation (diffraction-limited focus) and depletion (vortex beam) lasers at the output of the respective different fibres as mentioned and labelled in {\bf b-g}. The inset in {\bf k} shows the location of the generated beams in respect to the fibre core (circle line). In all cases, the beams were created by controlling a single circular input polarisation and by employing phase-only modulation. Scale bar: \SI{10}{\micro\metre}.
}
\label{fig1}
\end{figure*}

\begin{figure*}[ht!] 
   \centering
   \includegraphics[width=\textwidth]{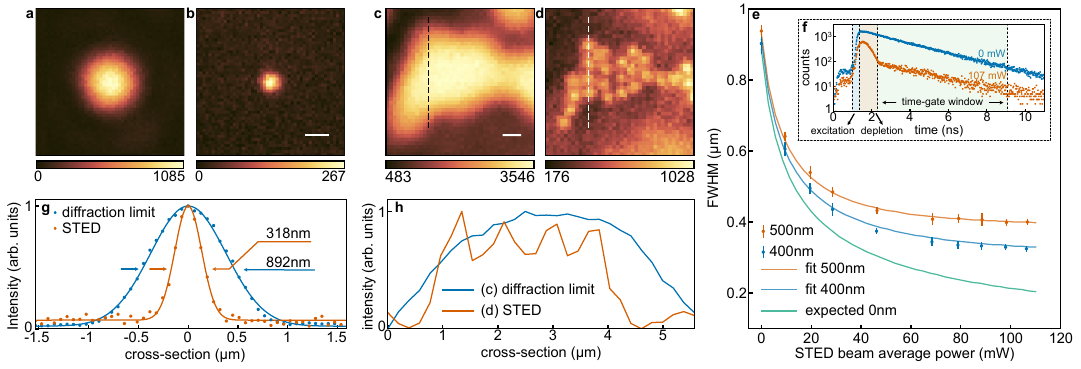} 
   \caption{ {\bf Performance of STED microscopy using pulsed lasers through an endcap fibre probe.} {\bf a}, Diffraction-limited image and {\bf b}, STED image of a \SI{400}{\nano\metre} fluorescent nanosphere. {\bf c}, Diffraction-limited image and {\bf d}, STED image of a \SI{400}{\nano\metre} fluorescent nanosphere cluster. {\bf e}, FWHM dependence on the average STED beam power determined from imaging \SI{500}{\nano\metre} and \SI{400}{\nano\metre} fluorescent nanospheres. Each datapoint and corresponding error bar is the result from an average between 10 images of distinct fluorescent nanospheres. The curve for \SI{0}{\nano\metre} outlines the projection of the fitting model to an infinitely small fluorescent centre, indicating an effective point-spread function FWHM close to \SI{200}{\nano\metre}. {\bf f}, Example of a time-correlated single-photon counting histogram from a \SI{500}{\nano\metre} fluorescent nanosphere measured at two indicated STED beam powers ($0$ and \SI{107}{\milli\watt}). The time-gate detection window starts after the excitation and depletion events, avoiding counting photons not yet depleted. {\bf g} Normalised cross-section of {\bf a} and {\bf b} (vertical and horizontal average) demonstrating a resolution improvement of about 3-fold over the diffraction-limit. {\bf h}, Normalised cross-section corresponding to the dashed lines in {\bf c} and {\bf d}, revealing 4 resolved spheres in the STED image (limited by the sphere size).
   Average excitation power in {\bf a-d}: \SI{2.4}{\micro\watt}, {\bf e}: \SI{2.4}{\micro\watt} for \SI{400}{\nano\metre} spheres and \SI{0.6}{\micro\watt} for \SI{500}{\nano\metre} spheres; STED beam average power in {\bf b, d}: \SI{107}{\milli\watt}. Dwell time: \SI{1}{\milli\second}. Scale bars: \SI{500}{\nano\metre}.
   } 
\label{fig2}
\end{figure*}

\begin{figure*}[ht!] 
   \centering
   \includegraphics[width=\textwidth]{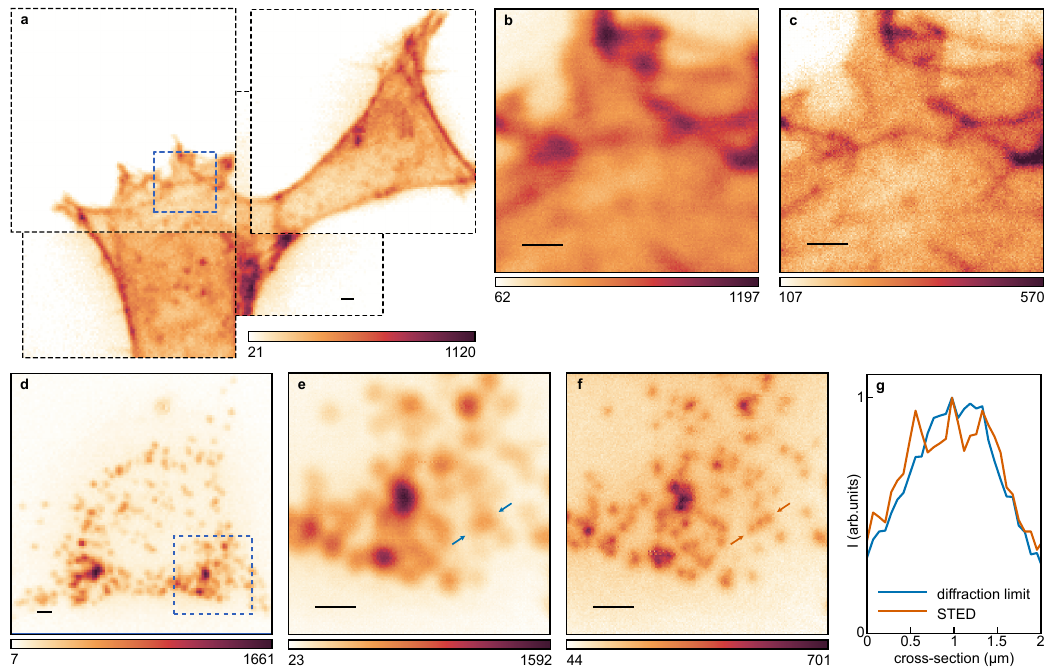} 
   \caption{ {\bf STED imaging of cells through an endcap fibre endoscope.} {\bf a}, Panel consisting of several diffraction-limited images of a HEK293 cell, in vitro, with actin stained with Alexa Fluor 488 Phalloidin. {\bf b}, Diffraction-limited image and {\bf c}, STED image of the dashed blue line region in {\bf a}. {\bf d}, Diffraction-limited image of peroxisomes in a HEK293 cell, in vitro, with sterol carrier protein 2 (SCP2) stained with enhanced green fluorescent protein (eGFP). {\bf e}, Diffraction-limited image and {\bf f}, STED image of the dashed blue line region in {\bf d}. {\bf g} Normalised cross-section of the arrow-marked regions in {\bf e} and {\bf f}, showing three distinguishable peroxisomes in the STED image, which are not resolved in the diffraction-limited image. Average excitation power in {\bf a-c}: \SI{1.5}{\micro\watt}, {\bf d-f}: \SI{1.0}{\micro\watt}; STED beam average power in {\bf c, f}: \SI{56.4}{\milli\watt}. Dwell time: \SI{1}{\milli\second}. Scale bars: \SI{2}{\micro\metre}.}
\label{fig3}
\end{figure*}

The dominant focus of this work is the realisation of a STED mechanism delivered through a hair-thin multimode fibre, namely the through-fibre synthesis of two distinct beams derived from pulsed laser light with different spectral content, precisely co-localised in space and time. From a practical perspective, achieving this requires the optical architecture to be as simple and mechanically stable as possible.
To limit the system complexity, we therefore restrict the optical design to a single, well-defined polarisation state for each pulsed beam. Allowing control over both polarisation states would necessitate independent wavefront control of two polarisation components, substantially increasing system complexity and the number of required optical elements, while imposing stringent long-term stability requirements.

When assessing fibre-type options for multimode fibre endoscopy, a standard step-index MMF is often considered an optimal choice. It maintains mode density across the facet, leading to uniform resolution across the field of view and near-perfect preservation of circular polarisation \cite{Gomes2022}.
A diffraction-limited focus at the MMF output is achieved through the constructive interference of all fibre modes (i.e. the full $k$-space) arriving in-phase at the target location. In a step-index MMF, light couples into and propagates through a myriad of different modes \cite{Snyder1983}, each travelling distinct optical path lengths. However, for a pulsed laser with a bandwidth of a few nanometres, such as the excitation laser, the optical path difference between lower- and higher-order modes in a few centimetre-long step-index MMF exceeds the coherence length of the laser. This results in loss of mutual coherence between modes (also known as modal coherence loss), where only certain modes (i.e. restricted $k$-space), which are still mutually coherence and capable of interfering with each other, can effectively be used to generate the focus. 
Under these conditions, the best achievable outcome is the formation of a quasi-Bessel beam, as only a small portion of the propagation constants can constructively interfere (similar to selective phase-conjugation in \cite{Morales-Delgado2015b}). 
The substantially increased excitation volume along all directions makes such output fields unsuitable for both 2-D and 3-D STED modalities (see Fig. \ref{fig1}b, k and {\bf Supplementary Note 1} for more details). 

	

When using graded-index MMFs instead, one must contend with the fact that these fibre types do not preserve any state of polarisation of the transported light \cite{Petrov1996, BoonzajerFlaes2018, Petrov2021}. Controlling a single polarisation state at their input results in delivered beams with a low power ratio (fraction of power contained in the desired beam relative to the uncontrolled background), as depicted in Fig. \ref{fig1}l. Here, the amount of uncontrolled light affects both, the excitation and depletion beams. The excitation produces undesired background fluorescence, while the depletion vortex beam suffers from increased intensity at the central vortex minimum, thereby compromising and limiting the STED performance \cite{Blom2017}. 

	

Our proposed solution is thus an endcap fibre consisting of both fibre types described above: an arbitrary long segment of step-index MMF spliced to a quarter-pitch graded-index MMF endcap. 
In step-index MMFs, light travels through a set of mutually orthogonal propagation-invariant modes \cite{Ploschner2015d}, which exhibit a pronounced conservation of the transverse component of the wave vector ($k_{z}$), i.e. the projection of the wave vector along the optical fibre axis \cite{Leite2021}.
In essence, in-coupling light with a defined angle, $\alpha$, will travel in the step-index segment predominantly as a single mode group (i.e. propagation-invariant modes with similar propagation constants) and emerge from the fibre as a cone with the same preserved angle, $\alpha$ (Fig. \ref{fig1}a). The higher the coupling angle, the larger the diameter of this concentric annular distribution of light at the fibre output. The role of the quarter-pitch graded-index endcap is to act as a lens, projecting the annular distribution of light leaving the step-index segment from the output fibre far field into the fibre facet plane. Additionally, the working distance of probe (i.e position of the focal plane) can be adjusted by tuning the length of the graded-index endcap. 

Focussing light on a desired position at the focal plane of the endcap probe solely requires to control the corresponding mode group that leads to the output annular light distribution overlapping with the focus location, thus suppressing intermodal dispersion. This is achieved by controlling the reciprocal annular distribution at the far field of the fibre input, as illustrated in Fig. \ref{fig1}a and in {\bf Supplementary Figure \ref{input-output-corr}}.

The endcap fibre probe is, therefore, a synergy between a step-index and a graded-index MMF, with benefits from both fibres types without their compromises. It not only enables focussing of pulsed light with suppressed associated intermodal dispersion (a feature typical of graded-index MMFs), but also presents circular polarisation conservation of step-index MMFs \cite{Gomes2022}. Under these conditions, the combined endcap probe properties allowed us to generate both pulsed STED beams (excitation focus and vortex beam depletion) with higher quality, as visible in Fig. \ref{fig1}d,g,j. In particular, vortex beams created at the focal plane of the endcap fibre have extinction ratios (ratio between the central minimum intensity of the vortex and its maximum intensity) above what is typically desired for STED microscopy (more details in {\bf Supplementary Note 2}).

Additionally, temporal broadening of the pulses due to intermodal dispersion should be minimised in the endcap fibre, as the focussed pulse at its output predominantly traveled within a mode group. This effect is, in general, strong in step-index MMFs. However, for the pulse durations (picosecond pulsed lasers) and fibre lengths used in this study, the impact of pulse broadening is minimal, as demonstrated in Fig. \ref{fig1}h-j. The endcap fibre produces output pulse durations similar to those of the graded-index MMF, while the step-index MMF shows a slight increase in pulse duration. It worth noting that pulse broadening could be critical in other applications involving the delivery and focussing of ultrashort pulses, e.g. femtosecond pulses, for which the endcap fibre would offer an advantage over step-index MMFs.


\subsubsection*{Time-gated STED through endcap fibre endoscope}
The ability to focus and shape pulsed laser light through an endcap fibre endoscope enabled us to demonstrate STED microscopy with time-gated detection. The system geometry, introduced in {\bf Methods}, was used to empirically determine the light transport through the fibre, i.e. the TM, for both the excitation and depletion beams. The TM was processed as described in {\bf Methods} to obtain a set of holograms. When applied to the spatial light modulator (SLM), each hologram modulates both the excitation and depletion input beams. After propagating through the fibre, these beams form a diffraction-limited excitation focus and a vortex beam, respectively, at a specific position at the endcap output focal plane. Each hologram corresponds to a single scanning position. Sequentially displaying the set of holograms on the SLM therefore results in the excitation focus and vortex beam scanning across the region of interest at the endcap output focal plane.
The depletion vortex beams were created with an additional step of imprinting a vortex phase distribution to the TM outputs (see {\bf Methods}).
Since the TM measurement for both beams was performed using the same calibration module, the excitation and depletion beams could be generated coincidentally, by default, at the fibre output focal plane, if no mechanical drift occurred during the TM acquisition. In the event of non-overlapping excitation and depletion beams, the displacement between them can be determined and compensated by imaging a fluorescent nanosphere (more details in {\bf Supplementary Note 3}).
The STED image was obtained by sequentially scanning both beams across a desired region of the field of view, while collecting the remaining emitted fluorescence through the same fibre into a time-correlated single photon counting detector (see imaging section in {\bf Methods}).

We characterised the performance of STED microscopy through the endcap fibre endoscope using fluorescent nanospheres. Figure \ref{fig2}a and \ref{fig2}b show a diffraction-limited (without the depletion laser) and a STED microscopy image of a \SI{400}{\nano\metre} fluorescent sphere, respectively, recorded with \SI{64x64}{\nano\metre} pixel size and a dwell time of \SI{1}{\milli\second}. Their cross-section is displayed in Fig. \ref{fig2}g, demonstrating a resolution improvement in the STED image (limited by the sphere size) close to 3-fold over the diffraction-limited scenario. To further showcase the resolution improvement, we imaged a cluster of \SI{400}{\nano\metre} fluorescent spheres in both, diffraction-limited (Fig. \ref{fig2}c) and STED (Fig. \ref{fig2}d) regimes. While the nanospheres are undistinguishable in the diffraction-limited image (Fig. \ref{fig2}c), STED microscopy is able to resolve each nanosphere of the cluster, as demonstrated by the cross-section in Fig. \ref{fig2}h.

Figure \ref{fig2}e shows the performance of the STED microscope in terms of spatial resolution. For this, we imaged fluorescent nanospheres of 2 distinct sizes (\SI{400}{\nano\metre} and \SI{500}{\nano\metre}) through the fibre endoscope under different depletion laser powers. The sizes of the imaged nanospheres were quantified by azimuthally averaging and fitting the respective full width at half maximum (FWHM) of the imaged intensity profile (more details in {\bf Methods}).
Each data point corresponds to the average of 10 distinct spheres (i.e. each sphere is only imaged once, avoiding the influence of photobleaching), with its respective standard deviation as error bar. 
The FWHM values reduce with increasing depletion power, reaching a plateau limited by the respective nanosphere FWHM, as expected for STED microscopy \cite{Hell1994}. Naturally, the FWHM results are influenced by the imaged sphere size, as it results from convolving the effective point spread function (PSF) with the sphere fluorescence distribution. A fitting model was used to access the system real resolution, taking into consideration the expression for STED lateral resolution and the sphere size (described in detail in {\bf Methods}). This model allows us to extrapolate the results for a hypothetical sphere size of \SI{0}{\nano\metre} (point source), therefore purely representing the FWHM of the effective PSF at each depletion laser power (Fig. \ref{fig2}e). The expected PSF FWHM at the maximum depletion laser power used (\SI{107}{\milli\watt}) is close to \SI{200}{\nano\metre}, suggesting that the fibre endoscope-delivered STED is capable of reaching resolution improvements over 4-fold compared to the diffraction-limit (about \SI{885}{\nano\metre} at \SI{0}{\milli\watt} depletion laser power).

All measurements were performed with time-gated detection to improve the imaging resolution and contrast \cite{Leutenegger2010,Vicidomini2013}. The time-gate corresponds to the temporal window during which the photons are detected. After excitation, the STED beam depletes the fluorescence around the excitation centre, resulting in a fast photon count reduction followed by the decay of the leftover photons after depletion, as depicted in fig. \ref{fig2}f. The time-gate window can be carefully selected to detect only the leftover photons, rejecting the initial photons while depletion is still occurring. The impact of the time-gate window can be observed in {\bf Supplementary Video 1}, where a STED image of a \SI{400}{\nano\metre} fluorescent sphere cluster iwas consecutively displayed for different starting positions of the detection time-gate.

\subsubsection*{Fibre endoscope STED in biological samples}
To demonstrate the applicability of the endcap fibre STED microscope in biological samples, we imaged human embryonic kidney (HEK) 293 cell line in vitro. To prevent the sample from being exposed to the depletion beam unnecessarily, an electro-optic modulator was introduced in the depletion beam pathway (see {\bf Methods}). This modulator acted as a gate, activating the depletion beam only when the detection was occurring. For this reason, the maximum available depletion power for cell imaging was only \SI{56.4}{\milli\watt}.  

Figure \ref{fig3}a consists of multiple diffraction-limited images of a HEK293 cell with Alexa Fluor 488 Phalloidin-labelled actin, arranged in a panel with \SI{384x384}{\nano\metre} pixel size and \SI{1}{\milli\second} recording dwell time. The region marked with a dashed blue line was re-imaged with increased sampling (\SI{64x64}{\nano\metre} pixel size) in fig. \ref{fig3}b and in STED modality with \SI{56.4}{\milli\watt} average depletion power in fig. \ref{fig3}c. The resolution improvement in the STED image (fig. \ref{fig3}c) is evident, with sharper and more defined actin filaments as opposed to the diffraction-limited scenario (fig. \ref{fig3}b).

Similarly, a HEK293 cell with enhanced green fluorescent protein (eGFP)-labelled sterol carrier protein 2 (SCP2) was imaged in diffraction-limit with \SI{384x384}{\nano\metre} pixel size and \SI{1}{\milli\second} recording dwell time, as shown in fig. \ref{fig3}d. eGFP-SCP2 is a peroxisomal matrix protein localised in peroxisomes, which are small intracellular organelles with a diameter size ranging from \SI{130}{\nano\metre} to \SI{650}{\nano\metre} \cite{Galiani2016}. Figure \ref{fig3}e and \ref{fig3}f display the diffraction-limited and STED image, respectively, of the blue dashed line region in fig. \ref{fig3}d. Both images consist of \SI{180x180}{} pixels with \SI{64x64}{\nano\metre} in size, recorded with a dwell time of \SI{1}{\milli\second}. The cross-section of the region indicated by the arrows in figs. \ref{fig3}e-f is represented in fig. \ref{fig3}g, revealing three individual peroxisomes in the STED image (fig. \ref{fig3}f), which are not visible in diffraction-limited case (fig. \ref{fig3}e).

\subsection*{Discussion}
We have demonstrated that the method of pulsed STED super-resolution microscopy can be introduced through a composite multimode optical fibre. The resulting instrument represents a hair-thin endoscopic device capable of ``funnelling'' this super-resolution imaging modality into large depths of sensitive structure in a uniquely atraumatic manner.

In the heart of this invention is the realisation that the composite (end-capped) fibre probe is capable of delivering pulsed laser light with suppressed modal dispersion mechanisms, whilst maintaining the remarkable purity of the generated structured light fields. This enabling technology can find potential applications beyond the scope of this work, such as in focussing ultrashort pulses through a fibre endoscope for multiphoton imaging, confocal microscopy imaging by spatially resolved signal collection and selective rejection of out-of-focus light \cite{Pikalek2025}, or even in significant acceleration of other imaging modalities.

The performance assessment based on super-resolving fluorescent nanospheres of different sizes indicated a resolution improvement over 4-fold relative to the diffraction-limit. At the maximum depletion power available, the estimated FWHM of the STED point-spread function is close to \SI{200}{\nano\metre}. This is limited by the NA levels of the commercially available graded-index fibres ($0.29$ for this study), which may significantly improve in the near future.

We showcased time-gated STED imaging performed in HEK293 cells with fluorescently-labelled actin and peroxisomes, validating the applicability of this holographic endoscopic technology to bio-imaging. Although 3D-STED modality was not explored in this work, wavefront shaping generation of the respective depletion beam through the fibre is possible (see {\bf Supplementary Figure \ref{axial_extent}}). The current imaging rate is limited by the slow liquid-crystal SLM, used to avoid dispersion in contrast to fast digital micromirror devices employed in standard holographic endoscopes. Such issue would be overcome with upcoming technologies involving fast MEMS modulators based on piston-like micromirror arrays, which would increase the scanning speed by 1 to 2 orders of magnitude \cite{Rocha2024}.

STED microscopy is one of few modalities of super-resolution compatible with holographic endoscopes, since it relies on raster scanning. By utilising the scanning of a single spot with a local intensity minimum, one could equally well transfer our current approach to recent super-resolution MINFLUX microscopy \cite{Balzarotti2017}. Evolving into other imaging modalities would require the need for an equivalent of a fibre inverse, such as an optical inverter based on multi-plane light conversion \cite{Kupianskyi2024}, to avoid the spatial scrambling of information inherent to multimode fibres.

Among the many potential applications of this technology, the field of neuroscience can readily benefit from its unique powers. Here, the demonstrated improvements in resolution would greatly assist in monitoring, classifying, and studying the plasticity of structural manifestations of inter-neuronal signalling connections (dendritic spines, axonal boutons) in deep brain structures within living animal models.
 
\section*{Acknowledgments}
The authors would like to acknowledge support from 
the European Research Council (724530),
the Ministry of Education, Youth and Sports of the Czech Republic
(CZ.02.1.01/0.0/0.0/15\_003/0000476),
the European Regional Development Fund (LM2018129),
the European Union's H2020-RIA (101016787),
The authors further greatly acknowledge financial support by the Deutsche Forschungsgemeinschaft (DFG, German Research Foundation; Germany's Excellence Strategy - EXC 2051 - Project-ID 390713860; project number 316213987 - SFB 1278; GRK M-M-M: GRK 2723/1 - 2023 - ID 44711651), 
the State of Thuringia (TMWWDG), 
and the Free State of Thuringia (TAB; Advanced Flu-Spec / 2020 FGZ: FGI 0031). 
Further, this work is integrated into the Leibniz Center for Photonics in Infection Research (LPI). The LPI initiated by Leibniz-IPHT, Leibniz-HKI, UKJ and FSU Jena is part of the BMBF national roadmap for research infrastructures.
Further, we would like to thank Yang Du and Ivo Leite for useful discussions and advice, and Jan Dellith for support with the electron microscope image assessment.

\section*{Data Availability}
The raw datasets from spatial (cross-section and axial extent) and temporal profile of the output beams, as well as all raw images of the fluorescent nanospheres and HEK293 cells obtained with the STED holographic endoscope have been deposited in the Zenodo repository \href{https://doi.org/10.5281/zenodo.14637567}{\color{blue}https://doi.org/10.5281/zenodo.14637567}. The raw data and processed data related to the STED resolution dependence on the average STED depletion beam power of the STED holographic endoscope can be found in \href{https://doi.org/10.5281/zenodo.14639005}{\color{blue}https://doi.org/10.5281/zenodo.14639005}.

\section*{Code Availability}
The python script to analyse the STED resolution dependence on the average STED depletion beam power of the STED holographic endoscope from Figure \ref{fig2}e, as well as the fitting model, can be assessed at the Zenodo repository \href{https://doi.org/10.5281/zenodo.14639163}{\color{blue}https://doi.org/10.5281/zenodo.14639163}. The python script to determine the extinction ratio of Vortex beams produced through the STED holographic endoscope based on HDR images can be found in \href{https://doi.org/10.5281/zenodo.14643188}{\color{blue}https://doi.org/10.5281/zenodo.14643188}.

\section*{Author Contributions}
T.Č. was responsible for the overall methodological conceptualisation. A.G and T.Č conceived the technological background. C.E. and T.Č. conceived the bioimaging application.
A.G. assembled the experimental system, with support and feedback from S.T. and T.Č..
A.G. and S.T. compiled the computer controlling software and interface.
M.S. and T. P. developed and manufactured the endcap fibre probes.
K.R. and C.E. prepared the HEK293 cell samples, including transfections and staining with the respective fluorescent markers.
A.G. performed all the imaging experiments.
A.G. and T.Č. wrote the scripts for data processing and analysed the data.
T.Č and C.E. secured funding and led the project.
A.G wrote the manuscript with contributions from all authors.

\section*{Competing Interests}
A.G., M.S., S.T., T. P., and T.Č. are authors of a related patent (DE102024202794B3 - Endoscopic detector system, composite optical fibre, endoscopic system and method for examining a sample).
S.T., and T. Č. are co-founders of the startup DeepEn (\url{https://deepen-imaging.com}). 
The other authors declare no competing interests.

\section*{METHODS}
\subsection*{Experimental system}
The  complete  layout  of  our  experimental system is  depicted in {\bf Supplementary  Figure \ref{setup}.} A collimated \SI{592}{\nano\metre} high power picosecond depletion laser (Katana 06 HP, NKT) operating at \SI{20}{\mega\hertz} was gated by a resonant electro-optic modulator EOM (AM2B-VIS\_0.1, QUBIG), together with the polarising beamsplitter PBS1. When the gate was active, a DAQ card (DAQ BNC-2110, National Instruments) generated an AC voltage signal (\SI{125}{\kilo\hertz}, \SI{50}{\percent} duty cycle), which was amplified by a wide band amplifier (TOE 7607, TOELLNER) and applied to the EOM. This resulted in a beam polarisation rotation of \SI{90}{\degree}, allowing the beam to be transmitted through PBS1. When the gate was inactive, the PBS1 reflected the beam into a beam block. Note that the results in fig. \ref{fig2} were obtained without EOM. The computer-controlled half-wave plate H1 and polarising beamsplitter PBS2 allowed controlling the total amount of power propagating in the optical system, being the excess power sent to a beam block. Lenses L1 ($f =$ \SI{40}{\milli\metre}) and L2 ($f =$ \SI{250}{\milli\metre}) form a telescope, magnifying the original beam size to match the \SI{600x600}{px} area of interest (AOI) of the spatial light modulator SLM (HSP1920-500-1200, Meadowlark). The half-wave plate H2 aligned the beam polarisation to match the polarisation axis of the SLM. The beam was transmitted through a dichroic mirror DM1 (Longpass Dichroic Mirror, \SI{505}{\nano\metre} Cut-On, Thorlabs) and divided into a signal beam (\SI{90}{\percent}) and a reference beam (\SI{10}{\percent}) by the beamsplitter BS1. The signal beam was reflected on the dielectric mirror M1 towards the left half of the SLM.

The computer-controlled half-wave plate H3 and polarising beamsplitter PBS3 controlled the total amount of power in the optical system of a collimated \SI{485}{\nano\metre} picosecond excitation laser (LDH-IB-485-B, PicoQuant) operating at \SI{20}{\mega\hertz}. The telescope consisting of lenses L3 ($f =$ \SI{35}{\milli\metre}) and L4 ($f =$ \SI{45}{\milli\metre}), together with the aspheric lens L5 ($f =$ \SI{8}{\milli\metre}, NA $= 0.5$), resized and focussed the laser beam into a single-mode polarisation-maintaining fibre PM-fibre, here employed as a gaussian spatial filter. The output of the PM-fibre was expanded and collimated by lens L6 ($f =$ \SI{80}{\milli\metre}). The half-wave plate H4 aligned the excitation beam polarisation with the SLM. The excitation beam was merged with the depletion beam by reflecting on the dichroic mirror DM1 and sent towards the right half of the SLM by the dielectric mirror M7.

The SLM was split in half to control simultaneously both, the excitation and depletion beams. The light from both beams reflected off the SLM travelled through lens L7 ($f =$ \SI{250}{\milli\metre}) and the dichroic mirror DM2 (STED Laser Beamsplitter zt 488 RDC, AHF Analysentechnick AG) separated the excitation and depletion beams into different paths (depletion was transmitted through and excitation was reflected off DM2). The SLM plane was imaged onto the MMF input facet using two telescopes: L7 ($f =$ \SI{250}{\milli\metre}) and L8 $=$ L10 ($f =$ \SI{45}{\milli\metre}), followed by L9 $=$ L11 ($f =$ \SI{150}{\milli\metre}) and objective Obj1 (10x, Olympus Plan Fluorite), for the excitation and depletion paths, respectively. These two optical paths were merged together using the dichroic mirror DM3 (Shortpass Beamsplitter T 565 SPXR, AHF Analysentechnick AG). The SLM was employed in the off-axis regime, where the first diffraction order of the holograms displayed at the SLM was aligned with the optical axis using the dielectric mirrors M8 and M9 for the excitation path, and the dielectric mirrors M10 and M11 for the depletion path. In this configuration, all power in the first diffraction order was coupled into a particular $k$-vector at the input fibre facet (i.e. into a plane wave with a specific input angle $\alpha$) \cite{Leite2021}. The polarisation state of the excitation and depletion beams was converted to circular before the objective Obj1 using the half-wave plate H5 and quarter-wave plate Q1, as well as the half-wave plate H6 and quarter-wave plate Q2, respectively, since circular polarisation is well-conserved when propagating through step-index MMF segments (which is the main component of the endcap fibre probe).

The samples were mounted on a microscope slide holder connected to a 3-axis stage, remotely controlled using 3 picomotors for precise navigation of the sample in all directions with respect to the fibre endoscope. The whole assembly was carried by a robust high-load vertical translation stage, which enabled a coarse vertical positioning of the sample in relation to the fibre endoscope. The fibre output focal plane or the sample were imaged onto a camera chip CAM (CMOS, Basler) using the long working distance objective Obj2 (50x, Mitutoyo Apochromatic) and tube lens L14 ($f =$ \SI{100}{\milli\metre}). The working distance of Obj2 (\SI{13}{\milli\metre}) allowed enough space to place and manipulate the sample between the endoscope output and the objective.

The collimated reference beam (\SI{10}{\percent} coming from the beamsplitter BS1, for both excitation and depletion, was magnified with the lens pair L15 ($f =$ \SI{50}{\milli\metre}) and L16 ($f =$ \SI{100}{\milli\metre}) in a $4f$ configuration, and merged with the signal beam at the beamsplitter BS3 (50/50) onto the camera chip CAM, for interferometric measurement of the TM. The optical path of the reference arm could be adjusted to match that of the signal arm using an optical delay line composed of the dielectric mirrors M14, M15, M16, and M17, being M15 and M16 on a translation stage with \SI{100}{\milli\metre} travel range. When calibration was not occurring, a flipping mirror inserted between lenses L15 and L16 diverted the reference beam towards a power meter, monitoring the current power in the optical system.

The fluorescence photons from the sample were collected through the same fibre endoscope and travelled back in the system through the excitation arm. The dichroic mirror DM2 separated the fluorescent signal from the excitation and depletion beams, and lens L13 ($f =$ \SI{200}{\milli\metre}) focussed the fluorescence signal on a hybrid photomultiplier detector (PMA Hybrid 40, PicoQuant). The set of filters F (notch filter NF488-15, Thorlabs; 2x notch filter NF594-23, Thorlabs; BrightLine fluorescence filter 531/40, Semrock) spectrally separated the fluorescent photons from residual excitation or depletion laser light before the detector.
A white LED (warm white, \SI{1}{\ampere}, \SI{500}{\milli\watt}, Thorlabs), together with a ground glass diffuser D (220 grit, Thorlabs), was incorporated in the optical system after lens L12 ($f =$ \SI{40}{\milli\metre}) by fresnel reflection on a precision cover glass G (\SI{170}{\micro\metre} thick) placed at \SI{45}{\degree}, providing a pseudo-Köhler illumination of the sample after propagating through the fibre endoscope.

It is challenging to navigate through the sample while imaging, since the scanning rate (point-by-point) is dictated by the rather low refresh rate of the liquid-crystal SLM (\SI{100}{\hertz}). Hence, we employed a digital micromirror device DMD (ViALUX V-7001) with a \SI{488}{\nano\metre} CW laser (Coherent Sapphire SF 488) to allow a fast pre-scan of the field of view, helping navigating through the sample and focussing in a region of interest. The CW laser was pre-coupled into 2 polarisation-maintaining fibres, one for the signal beam Laser 3, and the other for the reference beam REF. Laser 3, with adjustable power split between the two fibres. The signal beam Laser 3 was collimated by lens L17 ($f =$ \SI{200}{\milli\metre}) and illuminated the DMD under an incident angle of $\sim$\SI{24}{\degree}. The first diffraction order of the holograms projected on the DMD was integrated into the excitation beam pathway through \SI{10}{\percent} reflection on the non-polarising beamsplitter BS2 (10/90), with help of the dielectric mirror M18. The DMD was used in the off-axis regime to allow phase-only modulation of light \cite{Turtaev2017}. The reference signal was merged onto the calibration module by reflection on the non-polarising beamsplitter BS4 (50/50) and collimated onto the camera chip CAM by lens L14 ($f =$ \SI{100}{\milli\metre}). 

The optical and mechanical components were assembled to minimise mechanical drifts. The optical system was placed on an optical table with active dumping of vibrations and the room temperature has been controlled. Mechanical drifts during calibration can result in an incorrect determination of the TM and even lead to non-overlapping excitation and depletion beams (see {\bf Supplementary Note 3}).

The whole system was controlled with a computer station, also used to process the TMs and calculate the holograms. The TM acquisition was performed using a home-built computer interface in National Instruments LabVIEW\textsuperscript{\texttrademark}. All TM processing and hologram calculations were executed in python using GPU-accelerated computing with CUDA. The computer runs on Windows 10, with an AMD Ryzen 7 3700X 8-core processor, 128 GB RAM, 1 TB SSD, and an NVIDIA RTX A6000 (47.5 GB dedicated memory) GPU.

\subsection*{Triggers and communication protocols}
The communication and triggers between the different components were assembled in 4 different configurations, depending on each of the following procedures: calibration using SLM, calibration using DMD, imaging using SLM, and imaging using DMD. 
In the calibration procedure using the SLM, the In/Out triggers from the SLM were directly connected to the In/Out triggers of the camera. Analogously, during calibration using the DMD, the In/Out triggers of the DMD were directly connected to the In/Out triggers of the camera.

While imaging with the SLM, the hybrid photomultiplier detector output was connected to a time-correlated single photon counting (TCSPC) unit (PicoHarp 300, PicoQuant). This unit received the synchronisation signal from the excitation laser, a marker signal, and a gate signal. The SLM trigger Out was connected to a pulse delay generator (TOMBAK, Aerodiode), which converted the trigger into a fast pulse (\SI{100}{\micro\second}) to be used as a marker signal. The TCSPC unit inserted this marker in the photon counting datastream, allowing to locate in the data when a new scanning position has started (i.e. new hologram on the SLM). A second pulse delay generator, in parallel with the first one, generated a gate signal of \SI{1}{\milli\second} duration with a delay of \SI{8}{\milli\second} upon receiving the SLM trigger Out. This gate signal was split into 2: one linked to the TCSPC unit, and the second used to Gate the depletion laser. The liquid-crystals on the SLM require a certain amount of time to settle when displaying a new hologram. After assessment of this transient time, we concluded that a waiting time of \SI{8}{\milli\second} safely guaranteed that the hologram was settled, and imaging of the sample for that specific position could start. Hence, the Gate signal enforced the photon counting and depletion laser to wait for this process. The gate signal duration was set to \SI{1}{\milli\second} in all measurements, corresponding to the imaging dwell time. Nevertheless, it could be adjusted on demand, if necessary. The signal to Gate the depletion laser was connected to a DAQ card used to activate the electro-optic modulator (see experimental system in {\bf Methods}). The depletion laser acted as a Master oscillator (\SI{20}{\mega\hertz} repetition rate). Its repetition rate signal was connected to a third pulse delay generator, which delayed the repetition rate signal and fed it to the excitation laser. This way, the excitation laser runned at the same repetition rate as the depletion laser, and the delay between both lasers could be adjusted on demand using the pulse delay generator.

For fast pre-scan navigation with the DMD, the analog output voltage of the hybrid photomultiplier was measured with the DAQ card (BNC-2110, National Instruments). The DMD trigger Out was also connected to the same DAQ card, triggering it to read N-samples of the detector voltage values, which were then used to build the image.

\subsection*{Calibration procedures}
The TM for the excitation and depletion beams was empirically measured using a calibration procedure by means of phase-step interferometry \cite{Gomes2022}. Previously, wavefront correction (WFC) was performed to measure the aberrations in both optical paths (excitation and depletion beam), between the SLM and the back aperture of objective Obj1, following the same technique as described in \cite{Cizmar2010b, Gomes2022}. The input basis of the TM was a set of wavefront-corrected diffraction-limited foci (\SI{107x107}{} points) at the Fourier plane of the input fibre facet, corresponding to in-coupling plane waves with different orientations covering the whole NA of the input fibre facet. The output basis consisted of a square grid of \SI{480x480}{} camera pixels, where the endcap probe focal plane was imaged onto. The TM was successively measured for the excitation and depletion beams, and corrected for phase drifts of the reference arm with respect to the signal arm \cite{Leite2018a}. The delay line in the reference arm (see experimental system in {\bf Methods}) was adjusted to have maximum constructive interference with the signal beam on the camera. Between TM measurements, the delay line was re-adjusted to match the arm length for maximum interference with the following beam to be calibrated. 

Given the low coherence length of the excitation laser, and to facilitate the calibration procedure without constantly needing to act on the delay line to match in real time the optical paths, the excitation laser calibration was performed with the laser in CW operation. The acquired TM for CW excitation was then employed in the pulsed regime. Since in CW operation the bandwidth of the laser (around \SI{1.3}{\nano\metre}) matches with the wavelength region containing most of the optical power in the pulse regime, the created diffraction-limited focus of pulsed excitation using the TM measured in CW operation still led to good focussing with low background (see fig. \ref{fig1}m).

As presented in the results, for the case of a pure step-index fibre, the coherence length of the laser is smaller than the optical path difference between the higher and lower order modes in the fibre. Therefore, one must choose around which mode group the reference arm is adjusted to maximise the interference. In case of the bessel beam showcased in fig. \ref{fig1}b and k, we opted to tune the reference arm to maximise the interference with intermediate mode groups. The result of adjusting the reference arm to interfere with lower or higher order modes also leads to bessel beams, but with different axial extents, as can be seen in {\bf Supplementary Figure \ref{ref_tuning}}. 

The calibration procedure was performed with the output fibre facet immersed in immersion fluid (Immersol W, n=1.334 (\SI{23}{\celsius}), Carl Zeiss, Germany) over a precision cover glass for sphere imaging, or in phosphate-buffered saline (PBS) solution on a \#1.5 \micro -dish (\SI{35}{\milli\metre}, high glass bottom, ibidi GmbH, Germany) for cell imaging. Performing the calibration in the same medium and conditions as for imaging minimises possible aberrations.

The TM measurement with the DMD and the additional CW laser was executed in a similar manner as for the SLM. The aberrations caused by the DMD were previously measured with the same WFC technique and compensated in every hologram displayed. Moreover, the DMD chip was actively temperature controlled, as described in \cite{Rudolf2021}. A set of binary amplitude gratings based on the Lee hologram \cite{Lee1979} created a similar input basis of wavefront-corrected diffraction-limited foci at the Fourier plane of the input fibre facet. The reference beam did not require to be adjusted with the delay line, since the laser was single-frequency (long coherence length).

\subsection*{Transmission matrix processing}
Once acquired, the TMs were processed in various ways to generate a sequence of holograms that produce excitation foci, depletion vortex beams, or even shift the focal plane of the beams.

\subsubsection*{Output resampling}
The magnification of the telescope imaging the fibre output focal plane onto the camera led to an effective pixel size of \SI{192}{\nano\metre} (sampling of the TM outputs). In fig. \ref{fig3}a and d, the TM outputs were decimated by a factor of 2, resulting in a sampling of \SI{384}{\nano\metre}. To image the cluster of \SI{400}{\nano\metre} fluorescent spheres in fig. \ref{fig2}c and d, the TM outputs were upsampled to twice the original size.
The upsampling was performed by extending the TM output dimension through zero-padding in the Fourier space. The number of outputs were extended to twice the size, leading to a sampling of \SI{96}{\nano\metre}.
For assessing the STED resolution performance, in fig. \ref{fig2}a, b, e, and g, as well as for cell imaging in fig. \ref{fig3}b, c, e, and f, the TM output dimension was upsampled to 3-times its size. The final sampling for these cases was \SI{64}{\nano\metre}. 

\subsubsection*{Output far-field shift correction}
Creating foci across the endcap output focal plane involves focussing the respective plane waves through the quasi-quarter pitch graded-index fibre segment. Under these conditions, creating a focus positioned away from the centre of the focal plane will induce a respective shift of its far-field (angular spectrum), as exemplified in {\bf Supplementary Figure \ref{phase_ramp}g}. 
In other words, the generated focus has a reduced effective NA (far-field projection) and carries a phase ramp distribution (see {\bf Supplementary Figure \ref{phase_ramp}b}), resulting from the reciprocal far-field shift. 
Such shift has implications in the generation of the vortex beams, since the required vortex phase mask must be centred with respect to the output far-field projection (see next subsection). 
To correct for the output far-field dislocation, the desired outputs were computationally generated as diffraction-limited foci by multiplying the respective conjugated input field with the transposed TM. Then, the phase ramp for each focus (see {\bf Supplementary Figure \ref{phase_ramp}b}) was obtained through a linear fit of the phase distribution around the central region of the foci, respectively in both orthogonal directions (X and Y). The slope of the linear fit (i.e. phase ramp) increases as a function of the focus offset position in relation to the centre of the focal plane (see {\bf Supplementary Figure \ref{phase_ramp}c and d}). At last, the desired TM outputs were multiplied with their respective reciprocal phase ramp, therefore countering the far-field shift.

\subsubsection*{Vortex beam generation}
The depletion vortex beams were generated by imposing a vortex phase distribution to the TM outputs. The vortex phase distribution ($v$) is expressed as:
\begin{equation}
\label{vortex_mask}
v = exp(il\phi),
\end{equation}
with $l=1$ and $\phi = arctan(k_x/k_y)$, where $k_x$ and $k_y$ are the X and Y components of the $k$-vector which, in the Fourier space, form an orthogonal basis and correspond to the angular spectrum representation.
First, the output far-field shift correction was applied to the desired TM outputs. Then, the output fields were projected into the angular spectrum representation, through a Fourier transform, and multiplied with the vortex phase distribution from eq. \ref{vortex_mask}. The new output fields (vortex beams) were converted back into the spatial coordinates through an inverse Fourier transform. These were then multiplied with the transposed TM to obtain the respective input fields.
It is vital to ensure the vortex phase mask is centred with respect to the output far-field. An off-axis vortex phase distribution leads to asymmetric vortex beams, with a higher intensity at the central singularity and a crescent intensity distribution (see {\bf Supplementary Figure \ref{phase_ramp}e}).

\subsubsection*{Refocussing}
The axial extent of the output beams in fig. \ref{fig1}b-g, as well as in {\bf supplementary fig. \ref{axial_extent} and  \ref{ref_tuning}}, was obtained through a series of camera images, for which the output beam focal plane was progressively shifted from \SI{-50}{\micro\metre} to \SI{200}{\micro\metre} in relation to the calibrated output focal plane, in steps of \SI{1}{\micro\metre}. This was performed by refocussing the output beams using an appropriate spherical phase mask calculated based on free-space propagation \cite{Stiburek2023}. The procedure is the same as for the vortex beam generation, where the TM output fields were projected into the angular spectrum representation, through a Fourier transform, and multiplied with the respective spherical phase distribution (D) described as:
\begin{equation}
D = e^{-i.k_z.z},
\label{defocus_mask}
\end{equation}
where $z$ is the defocus distance and $k_z$ is a grid in the Fourier space expressing the Z component of the $k$-vector, obtained as $k_z = \sqrt{k^2-{k_x}^2-{k_y}^2}$ (see vortex beam generation in {\bf Methods}).

\subsubsection*{Hologram generation}
Previously, the TM was measured in the input basis of diffraction-limited foci (first diffraction order of the SLM hologram) at the Fourier plane of the input fibre facet. 
Therefore, a complex input field required to generate a particular output is also expressed in the same input basis. A 2D Fourier transform of this complex field allowed to obtain the corresponding field at the SLM plane. The SLM WFC, as well as the grating with the respective carrier frequency, were subsequently applied to the input field at the SLM plane. Finally, the SLM hologram was calculated by converting the phase of the resultant input field (phase-only modulation) into the corresponding SLM grayscale values. The whole procedure was performed for both excitation and depletion TMs. The final hologram comprised both an excitation and a depletion hologram, each occupying one half of the final hologram.
A similar procedure was used to generate the DMD holograms. The 2D Fourier transform of the complex input field was multiplied with the DMD WFC, as well as with the grating with respective carrier frequency, followed by binarisation of the resultant field phase distribution using Lee holograms \cite{Lee1979}.

\subsection*{Imaging}
After acquiring the TMs, processing, and generating a sequence of appropriated holograms to scan an area of interest (for both the SLM and DMD), the quality and overlapping of the excitation and depletion beams was controlled before imaging using the camera in the calibration arm.
The sample was then placed in the sample holder and navigated towards the fibre endoscope with the remote-controlled 3-axis sample stage. First, the sample was illuminated through the fibre with the white LED and imaged onto the camera chip. This helped to navigate and pre-position the sample with respect to the endoscope. Note that due to the previous calibration procedure, the camera is currently imaging the focal plane of the endcap fibre endoscope.
Afterwards, the sample was imaged using the DMD, which enabled to target a region of interest and adjust the focus (i.e. the sample position in relation to the focal plane). The same region of interest was then imaged using the SLM with diffraction-limited excitation and coarse sampling (\SI{384}{\nano\metre} pixel size).
A smaller area of interest was subsequently selected and re-imaged with the excitation beam but with finer sampling (\SI{64}{\nano\metre} pixel size).
At last, the same region was re-imaged with both excitation and depletion beams (STED) with the same sampling.  An example of the imaging procedure (DMD fast pre-scan navigation to a region of interest and subsequent imaging with the SLM in pulsed diffraction-limited and STED regimes) is demonstrated in the {\bf Supplementary Video 2}. 

In the pulsed STED imaging modality, the delay between the excitation and depletion lasers was adjusted with a pulse delay generator (see triggers and communication protocols in {\bf Methods}). For our system in particular, the delay applied was \SI{75.6}{\nano\second}, so that each depletion pulse irradiates the sample hundreds of picosecond after the excitation pulse (about \SI{250}{\pico\second}). Naturally, such delay varies from system to system and depends on the optical path difference between the excitation and depletion arms.
The leftover fluorescence collected through the endcap fibre endoscope was detected by the hybrid photomultiplier detector and processed by the TCSPC module. The TCSPC module was set to count the detected photons with a time resolution of \SI{32}{\pico\second} (binning equal to 3). Moreover, a time offset of \SI{-9.6}{\nano\second} between the synchronisation signal from the excitation laser and the TCSPC acquisition time was introduced, plus an additional TCSPC acquisition time offset of \SI{41}{\nano\second}, thus leading to a time-gate window width of about \SI{4.5}{\nano\second}, starting after the depletion pulse. The TCSPC only registered the arriving fluorescence photons within this time-gate window. The TCSPC module and the depletion laser were gated with a \SI{1}{\milli\second}-wide pulse, generated with \SI{8}{\milli\second} delay in relation to the SLM trigger (see triggers and communication protocols in {\bf Methods}). Additionally, the SLM trigger was also introduced as a marker in the TCSPC data stream. An image was then build by reading the TCSPC module buffer, locating the position of the SLM markers, and the value of each pixel on the image resulted from the sum of all detected photons between two consecutive markers. The imaging procedure in the pulsed diffraction-limited imaging modality was the same, but with the depletion laser switched off.

\subsection*{Endcap fibre probe design and fabrication}
The endcap fibre probe was composed of a \SI{105}{\micro\metre}-diameter core, \SI{125}{\micro\metre}-diameter cladding and 0.22 NA step-index MMF (FG105UCA, Thorlabs) and a \SI{100}{\micro\metre}-diameter core, \SI{140}{\micro\metre}-diameter cladding and 0.29 NA gradient-index MMF (DrakaElite, Prysmian Group). The acrylate coating from both fibres was first stripped by approximately \SI{30}{\milli\metre} with a fibre stripping tool (FTS4, Thorlabs). Then, both fibres were right-angle cleaved on one end with a semi-automatic fibre cleaver (CT-105, Fujikura), and placed on V-grooves of a Large Diameter Splicing system (LDS 2.5, 3SAE Technologies, Inc). This splicing station allowed for precise alignment between both fibre terminations, given that they had different outer diameters. After splicing both fibres, the graded-index fibre was cleaved at \SI{386}{\micro\metre} from the splice region with an integrated piezoelectric cleaver, corresponding to the quarter-pitch length (approximately \SI{401}{\micro\metre}) minus \SI{15}{\micro\metre}. This set the focal plane of the endcap fibre probe to approximately \SI{15}{\micro\metre} in front of the output facet. The step-index fibre facet, opposite to the spliced near quarter-pitch graded-index fibre endcap, was cleaved to a desired length using the same semi-automatic fibre cleaver.
The total length of the fibre probe (step-index fibre segment plus graded-index endcap) was 25$\pm$\SI{2}{\milli\metre}. The input side of the probe was fixed into a \SI{10.5}{\milli\metre}-long zirconia ferrule (CF126, Thorlabs) using UV-curable glue (NOA 65, Norland). The remaining length of the fibre probe (approximately \SI{15}{\milli\metre}) could thus be used for insertion into the sample. A microscope image of an endcap fibre probe is depicted in {\bf Supplementary Figure \ref{endcap}.}

\subsection*{Sample preparation}
\subsubsection*{Fluorescent nanospheres}
We used \SI{400}{\nano\metre} and \SI{500}{\nano\metre} fluorescent nanospheres (w/v: \SI{1}{\percent}, Green-Fluo PS Microspheres with functional group -COOH, EPRUI Biotech Co. Ltd, China) diluted and dispersed in deionised water (\SI{2}{\micro\litre} spheres solution in \SI{1}{\milli\litre} deionised water). Droplets of this diluted solution were dispersed across a precision cover glass and let to evaporate, leaving fluorescent nanospheres scattered on the precision cover glass. Afterwards, a couple of drops from immersion fluid (Immersol W, n=1.334 (\SI{23}{\celsius}), Carl Zeiss, Germany) were placed on top of the dried nanospheres. The viscosity of the immersion fluid reduced the chance of nanospheres detaching from the glass surface and start brownian motion. 
The \SI{400}{\nano\metre} nanospheres imaged with a scanning electron microscope in {\bf Supplementary Figure \ref{SEM_sphere}} were prepared by diluting a part of the previously made solution by a factor of 10. Then a droplet of this solution was left to dry over an aluminium sample holder.

\subsubsection*{HEK293 cells}
HEK 293 cells (ATCC, USA) were maintained in a culture medium consisting of DMEM with \SI{4500}{\milli\gram} glucose/L, \SI{110}{\milli\gram} sodium pyruvate/L supplemented with \SI{10}{\percent} fetal calf serum, glutamine (\SI{2}{\milli\molar}) and penicillin-streptomycin (\SI{1}{\percent}). The cells were cultured at \SI{37}{\celsius}/\SI{8.5}{\percent} CO$_2$. Cells were grown on a \#1.5 \micro -dish (\SI{35}{\milli\metre}, high glass bottom, ibidi GmbH, Germany) and transfected with \SI{0.5}{\micro\gram} of eGFP-SCP2 \cite{Stanley2006} plasmid using Lipofectamine 2000 transfection reagent (Invitrogene, USA). The cells were fixed \SI{24}{\hour} after transfection with \SI{3}{\percent} formaldehyde for \SI{20}{\min} and stored in PBS afterwards. For staining actin phalloidin linked to Alexa Fluor 488 (Invitrogene, USA), the procedure according to the manufacturer's instructions was followed.

\subsection*{Data processing}
\subsubsection*{Nanosphere images}
The experimental data in fig. \ref{fig2}e were obtained from 100 images of \SI{400}{\nano\metre} and \SI{500}{\nano\metre} nanospheres, 10 at each depletion power. Each image was first subtracted by the respective background (i.e. a second image obtained by scanning the depletion laser, without excitation, collecting a small amount of fluorescent photons, which are excited by the depletion laser due to the high powers used \cite{Vicidomini2012}. 
Then, a centring algorithm based on least mean squares displaced the imaged sphere to the central position of the image. 
The centred images were azimuthally averaged by integration with circular masks of radii extending from the centre to the image periphery.
Afterwards, the azimuthally-averaged images were normalised to the maximum value and fitted with a gaussian distribution (G) expressed as:
\begin{equation}
\label{eq:gauss}
G = (a-y_0)e^{\left(\frac{-r^2}{2\sigma^2}\right)}+y_0,
\end{equation}
where $r$ is the radial position, $y_0$ is the offset, $a$ is the Gaussian amplitude subtracted by the offset (forcing the result for $r = 0$ to be 1, since the data was normalised to the maximum value), and $\sigma$ is the Gaussian standard deviation. At last, the FWHM value was obtained from the fitting parameter $\sigma$ as:
\begin{equation}
\label{eq:FWHM}
FWHM = 2\sqrt{2\log(2)}\left|\sigma\right|.
\end{equation}
An example of an azimuthally-averaged image fitted through equation \ref{eq:gauss} can be found in {\bf Supplementary Figure \ref{Gauss_Fit}}.

\subsubsection*{STED resolution fitting model}
The effective STED PSF can be approximated as a Gaussian with a reduced width \cite{Harke2008}, for which the FWHM can be described as:
\begin{equation}
FWHM = \frac{\lambda}{2NA\sqrt{1+I_{STED}/I_{sat}}},
\label{STED_resolution}
\end{equation}
where $\lambda$ is the excitation wavelength, $NA$ is the numerical aperture, $I_{STED}$ is the intensity of the STED beam, and $I_{sat}$ is the effective saturation intensity (intensity for which the probability of fluorescence emission is reduced by half).
The equation was incorporated into an optimisation algorithm to fit the experimental data from fig. \ref{fig2}e.
The algorithm minimised the difference between the measured FWHM, at different STED beam intensities, for both sphere sizes (\SI{400}{\nano\metre} and \SI{500}{\nano\metre}) and the respective simulated FWHM of the nanospheres.
The simulated FWHM corresponded to the convolution between a nanosphere and the STED PSF. In this particular case, the task could be simplified to a 1 dimensional problem, for which the STED PSF can be described as:
\begin{equation}
PSF(x) = exp\left[{-\left(\frac{2\sqrt{ln\left(2\right)}x}{FWHM}\right)^2}\right],
\label{Gaussian_PSF}
\end{equation}
with FWHM described by eq. \ref{STED_resolution}. Given that the fluorescence molecules are distributed throughout the entire volume of each nanosphere, the projection of this fluorescent volume along a line ($x$) crossing the centre of the sphere is defined as:
\begin{equation}
sphere(x) = real\left(\sqrt{d^2/4-x^2}\right),
\label{sphere_2D}
\end{equation}
being $d$ the diameter of the nanosphere. The fluorescent nanospheres were coated with a non-fluorescent functional group layer. As shown by the scanning electron microscopy image of the \SI{400}{\nano\metre} spheres in {\bf Supplementary Figure \ref{SEM_sphere}}, the sphere diameter indicated by the manufacturer included a non-fluorescent functional group layer with over \SI{10}{\nano\metre}-thickness. Therefore, the algorithm considered the nanosphere diameters to be \SI{25}{\nano\metre} smaller (\SI{12.5}{\nano\metre} in average for each side). The optimisation algorithm used the numerical aperture ($NA$) and effective saturation intensity ($I_{sat}$) as fitting parameters, finding the best match for which the simulated FWHM conforms to the experimental data. The fitting curves in fig. \ref{fig2}e reflect the converged values of $NA$ = 0.275 and $I_{sat}$ = \SI{6.16}{\milli\watt}. The algorithm is available in supplementary materials as a MATLAB code, together with the experimental data.

\bibliography{manuscriptV9.bib}

\clearpage
\onecolumn
\section*{\Large Supplementary Information}
\PaperTitle{Pulsed STED microscopy through holographic multimode fibre endoscope} 

\renewcommand\thefigure{S\arabic{figure}}    
\renewcommand\thesection{S\arabic{section}}    
\renewcommand\thesubsection{S\arabic{section}.\arabic{subsection}}    
\renewcommand\theequation{S\arabic{equation}}

\setcounter{figure}{0}    
\setcounter{equation}{0}    

\subsection*{Supplementary Note 1 - Axial extent comparison}
\label{Notes_axial_extent}
\begin{figure}[h!] 
\centering
\includegraphics[width=1\textwidth]{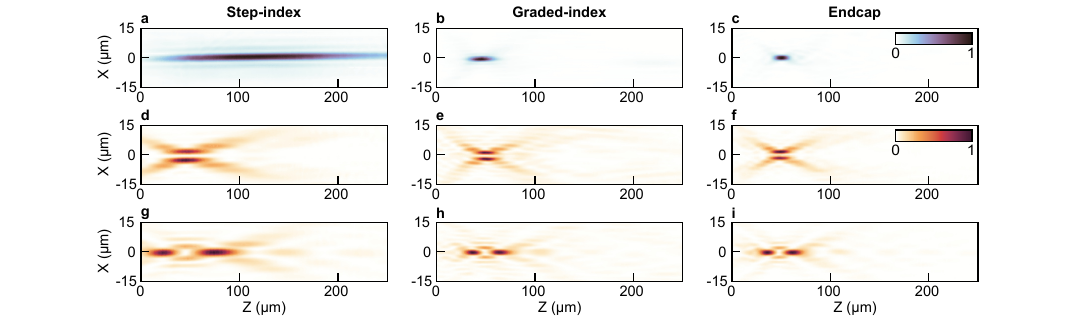} 
\caption{ {\bf Axial extent comparison between the pulsed excitation beam and pulsed depletion beam (vortex beam for 2D-STED and bottle beam for z-STED).} {\bf a,} Excitation beam, {\bf d,} depletion vortex beam, and {\bf g,} depletion bottle beam axial extent for a \SI{105}{\micro\metre} core and 0.22 NA step-index multimode fibre. {\bf b,} Excitation beam, {\bf e,} depletion vortex beam, and {\bf h,} depletion bottle beam axial extent for a \SI{100}{\micro\metre} core and 0.29 NA graded-index multimode fibre. {\bf c,} Excitation beam, {\bf f,} depletion vortex beam, and {\bf i,} depletion bottle beam axial extent for a \SI{100}{\micro\metre} core and 0.29 effective NA endcap multimode fibre. The calibrated focal plane is located at $Z = \SI{50}{\micro\metre}$.}
\label{axial_extent}
\end{figure}

Figure \ref{axial_extent} compares the axial extent of the excitation and depletion beams for three distinct fibres: \SI{105}{\micro\metre} core and 0.22 NA step-index multimode fibre, \SI{100}{\micro\metre} core and 0.29 NA graded-index multimode fibre, and \SI{100}{\micro\metre} core and 0.29 effective NA endcap multimode fibre. For both, the graded-index fibre and the endcap fibre, the axial extent of the excitation beam is within the axial extent of the depletion bottle beam, typically used in z-STED. However, the axial extent of the excitation laser in the step-index multimode fibre (a bessel beam) is larger than the depletion beam for z-STED. Therefore, the depletion bottle beam can never fully deplete out-of-focus light in a 3D sample.

The bottle beams were generated with a similar procedure as the vortex beams (see vortex beam generation in {\bf Methods}). Instead of imprinting a vortex phase distribution, a constant phase shift of $\pi$ in a central circular zone of the angular spectrum representation was added \cite{Cizmar2012}. For the case of a 0.22 NA step-index fibre, the radius of the $\pi$-shifted circular zone was set to \SI{0.10}{\radian}. For the graded-index, as well as for the endcap fibre, the radius of the $\pi$-shifted circular zone was \SI{0.13}{\radian}.

\begin{figure}[h!] 
\centering
\includegraphics[width=1\textwidth]{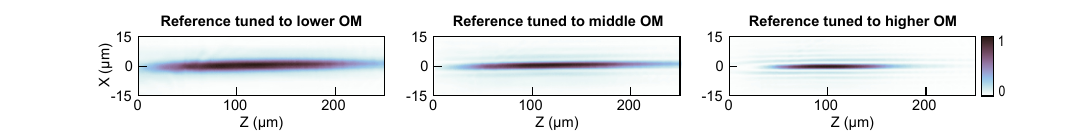} 
\caption{ {\bf Comparison of a pulsed excitation beam through a step-index multimode fibre}, in which the reference arm is tuned for maximum interference with the signal from {\bf a,} lower order modes (OM), {\bf b,} intermediate order modes, and {\bf c,} higher order modes.} 
\label{ref_tuning}
\end{figure}

Figure \ref{ref_tuning} compares the axial extent of the pulsed excitation beam through a step-index multimode fibre for 3 distinct cases, in which the reference arm is tuned for maximum interference with the signal from lower order modes, higher order modes, or modes in between the previous two (here referred to as intermediate order modes). As discussed in the manuscript, the coherence length of the excitation laser is smaller than the optical path difference between the higher and lower order modes. Hence, one must make a choice to which mode groups is the reference tuned to, providing maximum interference visibility. Consequently, the TM will only contain information restricted to those mode groups (restricted $k$-space). This leads to the formation of a bessel beam, when attempting to create a focus after the step-index fibre. In fig. \ref{fig1}b, we opted to display the intermediate case, where the reference arm was tuned to interfere with the intermediate order modes. As seen in fig. \ref{ref_tuning}, bessel beams with different axial extent or number of rings can result from the reference arm tuning conditions. Tuning the reference to lower order modes generate a bessel beam resulting from the interference of light leaving the fibre at low angles, thus having a large axial extent. On the opposite, tuning the reference to higher order modes leads to a bessel beam resulting from the interference of light leaving the fibre at high angles, reducing the axial extent of the bessel beam but increasing the number of rings.
 
\subsection*{Supplementary Note 2 - Vortex beam extinction ratio}
\begin{figure}[h!] 
\centering
\includegraphics[width=1\textwidth]{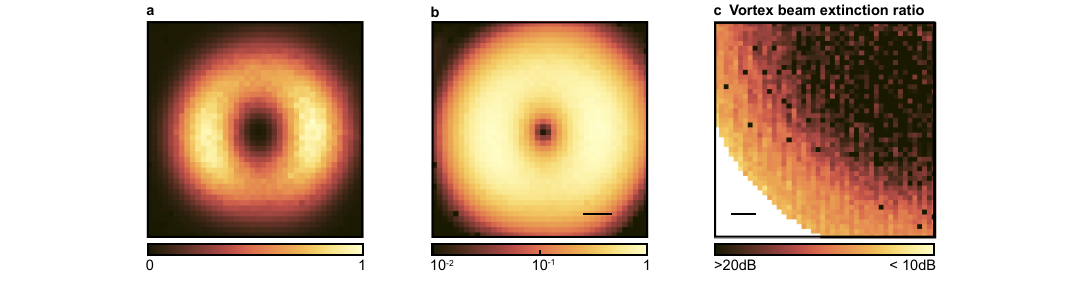} 
\caption{ {\bf Estimation of the vortex beams extinction ratio.} Example of a high dynamic range (HDR) image of a vortex beam in {\bf a,} linear scale and {\bf b,} logarithmic scale. {\bf c,} Extinction ratio of vortex beams generated within about $1/4$ of the field of view. The extinction ratio was estimated using the vortex beam minimum and maximum pixel values. The minimum extinction ratio desired for STED microscopy is \SI{13}{\decibel} \cite{Willig2006, Yan2019}. Scale bars: {\bf b}, \SI{500}{\nano\metre}, {\bf c}, \SI{5}{\micro\metre}.
}
\label{vortex_ratio}
\end{figure}

We have assessed the quality of vortex beams generated through the endcap multimode fibre probe, in particular the ratio between the minimum intensity in the vortex centre and its maximum intensity. The vortex beams must be imaged with large magnification in order to correctly sample the central minimum intensity. For this purpose, we changed the tube lens after Obj2 from \SI{100}{\milli\metre} to \SI{200}{\milli\metre}, allowing to record the vortex beams with a pixel size of \SI{96}{\nano\metre}. A series of vortex beams were generated across $1/4$ of the output field of view and recorded in a high dynamic range (HDR) image at 4 exposure times (\SI{69}{\micro\second}, \SI{690}{\micro\second}, \SI{6900}{\micro\second}, and \SI{69000}{\micro\second}). This allows to record both the high intensity levels, as well as the feeble low background. Figure \ref{vortex_ratio}a and \ref{vortex_ratio}b display an example of an HDR image of a vortex beam in linear and logarithmic scale, respectively. The extinction ratio ($R$) for the recorded vortex beams was estimated as:
\begin{equation}
\label{eq:extinction_ratio}
R = 10\log_{10}{(I_{max}/I_{min})}, 
\end{equation}
where $I_{max}$ and $I_{min}$ are the maximum and central minimum intensity of the vortex beam. The extinction ratio map within the measured $1/4$ of the field of view is shown in fig. \ref{vortex_ratio}c. The overall extinction ratio is greater than the desired minimum for STED microscopy of \SI{13}{\decibel} \cite{Willig2006, Yan2019}.

\subsection*{Supplementary Note 3 - Drift correction between excitation and depletion beam}
\begin{figure}[h!] 
\centering
\includegraphics[width=1\textwidth]{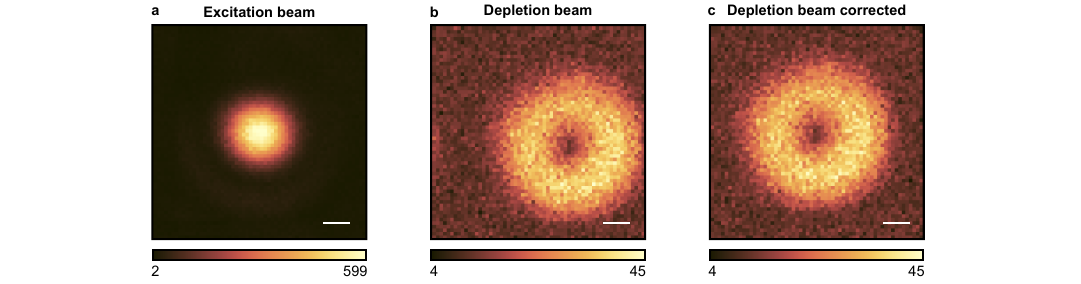} 
\caption{ {\bf Illustration of drift correction between excitation and depletion beams.} Average of 5 images of a centred \SI{500}{\nano\metre} fluorescent sphere {\bf a,} using the excitation beam and {\bf b,} only using the depletion beam. The high power of the depletion beam is sufficient to still induce some excitation on the fluorescent sphere, therefore allowing to visualise the depletion vortex beam alignment (result from convolution between the vortex beam and the sphere). The the vortex beam displacement to the centre of the image can be estimated and applied as a correction to all depletion beam outputs. {\bf c,} Image of the same fluorescent sphere using the depletion beam after displacement correction. The drift illustrated in this figure was exaggerated on purpose to showcase the correction procedure. Experimentally, the drift observed is typically ranging from sub-pixel to 2 pixels (i.e. smaller than \SI{150}{\nano\metre}).
Scale bars: \SI{500}{\nano\metre}.}
\label{shift_correction}
\end{figure}

\begin{figure}[h!] 
\centering
\includegraphics[width=1\textwidth]{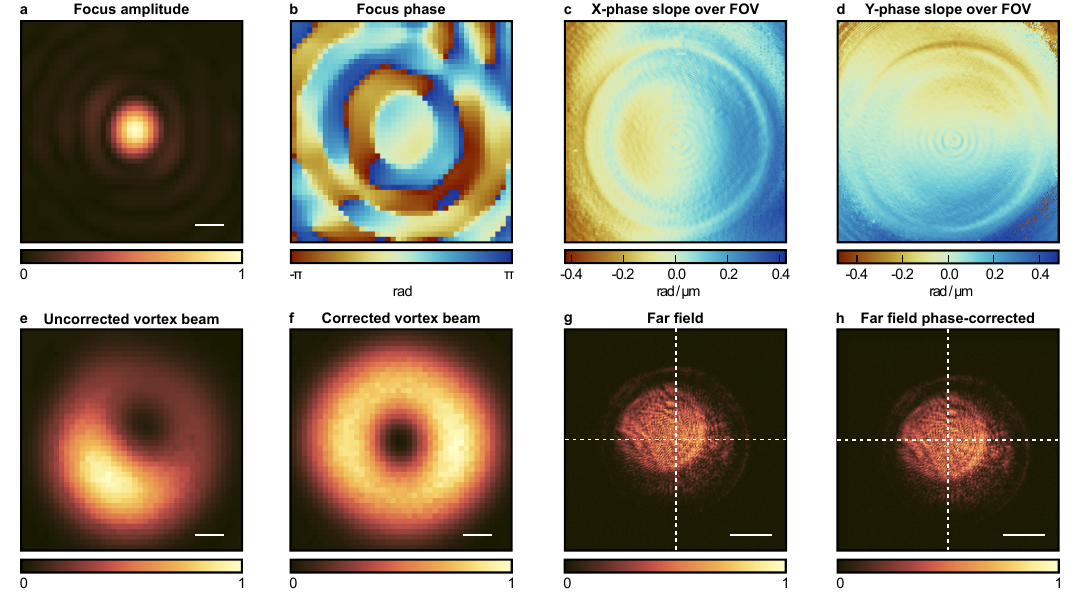} 
\caption{ {\bf Output farfield shift correction.} {\bf a,} Amplitude and {\bf b,} phase of a generated focus, radially displaced from the centre of the fibre endoscope output focal plane. The focus carries a phase ramp, which indicates a displacement of the corresponding focus farfield (angular spectrum). {\bf c, d,} Phase ramp slope, in X and Y directions, respectively, as a function of the focus position across the field of view (FOV). The phase ramp is stronger as the focus is generated farther from the centre. {\bf e,} Uncorrected and {\bf f,} phase-ramp corrected vortex beam. The displacement of the angular spectrum leads to asymmetric vortex beams, since the vortex phase distribution does not match with the centre of the effective angular spectrum. {\bf g,} Farfield amplitude of the focus in {\bf a} and {\bf b}. {\bf h,} Farfield amplitude after phase-ramp correction.
Scale bars: {\bf a, e, f}, \SI{500}{\nano\metre}, {\bf g, h}, \SI{\pi}{\radian\per\micro\metre}.}
\label{phase_ramp}
\end{figure}

\begin{figure}[h!] 
\centering
\includegraphics[width=0.5\textwidth]{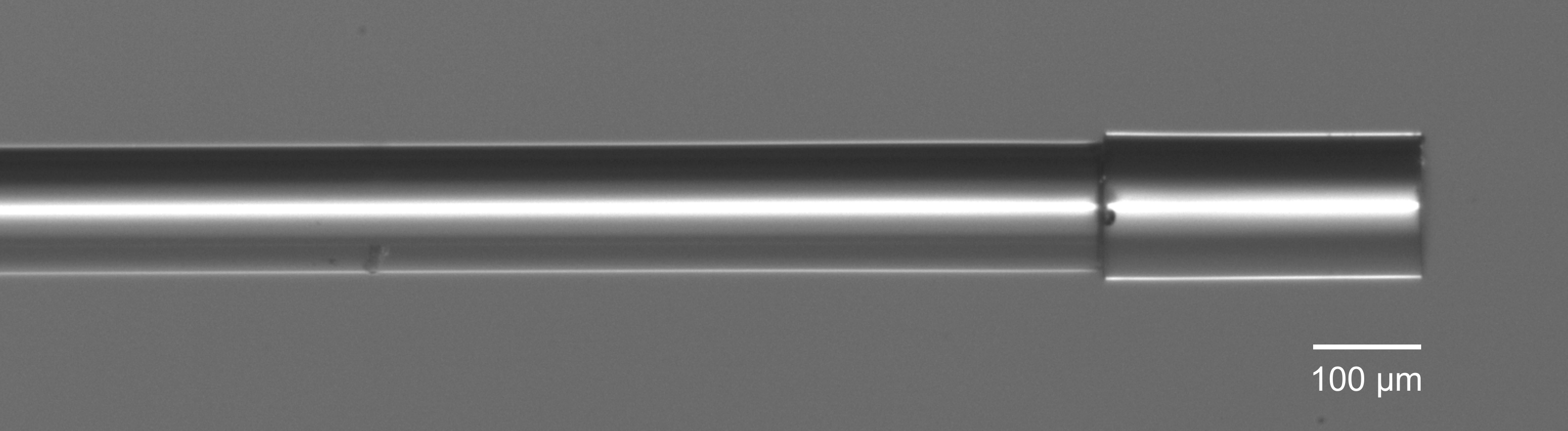} 
\caption{ {\bf Micrograph of a endcap multimode fibre probe.} The endcap is a quasi-quarter pitch length graded-index multimode fibre, spliced to a step-index multimode fibre segment.}
\label{endcap}
\end{figure}

\begin{figure}[h!] 
\centering
\includegraphics[width=0.5\textwidth]{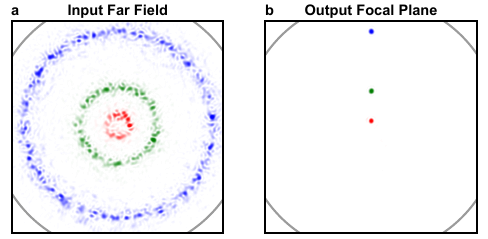} 
\caption{ {\bf Input-output correlation in the endcap multimode fibre probe.} {\bf a} Overlap at input farfield of the probe between the amplitudes of three distinct input fields. {\bf b} Overlap at the probe's output focal plane between the intensities of the corresponding foci, each resulting from the respective input field from {\bf a}. For visualisation, each displayed field is displayed in separate RGB colour channel. The grey circle represents the input numerical aperture in {\bf a} and the field of view at the focal plane in {\bf b}. Creating a focus at the probe's output focal plane mainly requires to control light within an annular region at the input farfield. Moreover, the radius of this annular distribution of light at the input far field is correlates with the radial position of the focus with respect to the optical axis of the fibre. The larger the radial position of the focus at the focal plane, the larger the radius of the respective annular light distribution at the input farfield. The plotted data is based on an experimentally acquired TM.}
\label{input-output-corr}
\end{figure}

\begin{figure}[h!] 
\centering
\includegraphics[width=1\textwidth]{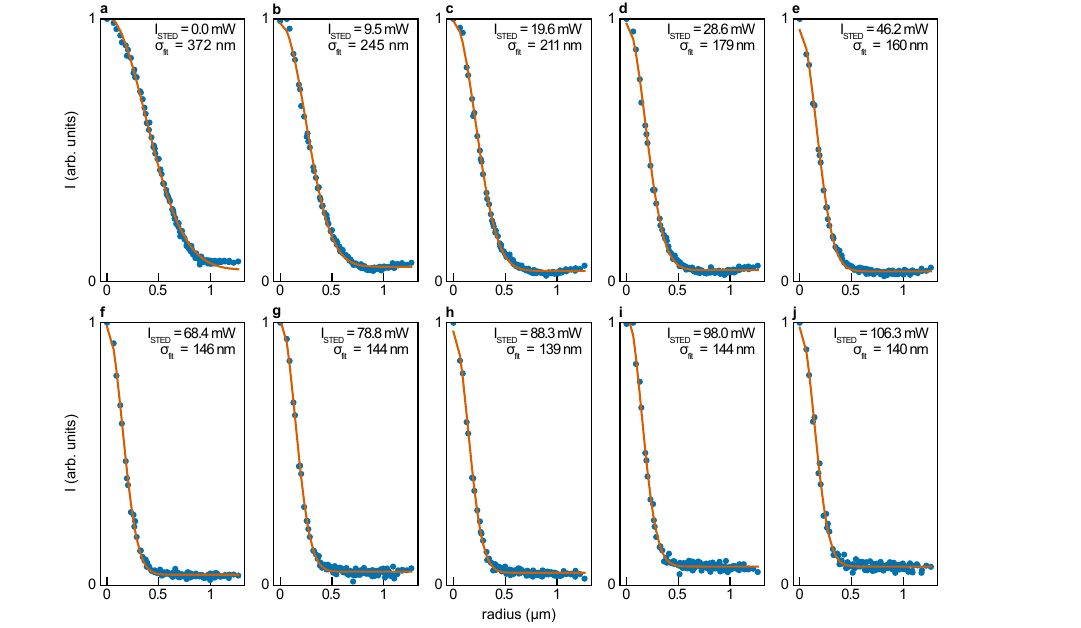} 
\caption{ {\bf Example of data processing used to assess the FWHM dependence on the average STED power.} The dataset (a-j) consists of distinct \SI{400}{\nano\metre} fluorescent spheres imaged at various depletion beam intensities ($I_{STED}$), azimuthally-averaged, and fitted through equation \ref{eq:gauss}. The fitting parameter corresponding to the gaussian standard deviation ($\sigma_{fit}$) is used to obtain the FWHM through equation \ref{eq:FWHM}.}
\label{Gauss_Fit}
\end{figure}

\begin{figure}[h!] 
\centering
\includegraphics[width=0.6\textwidth]{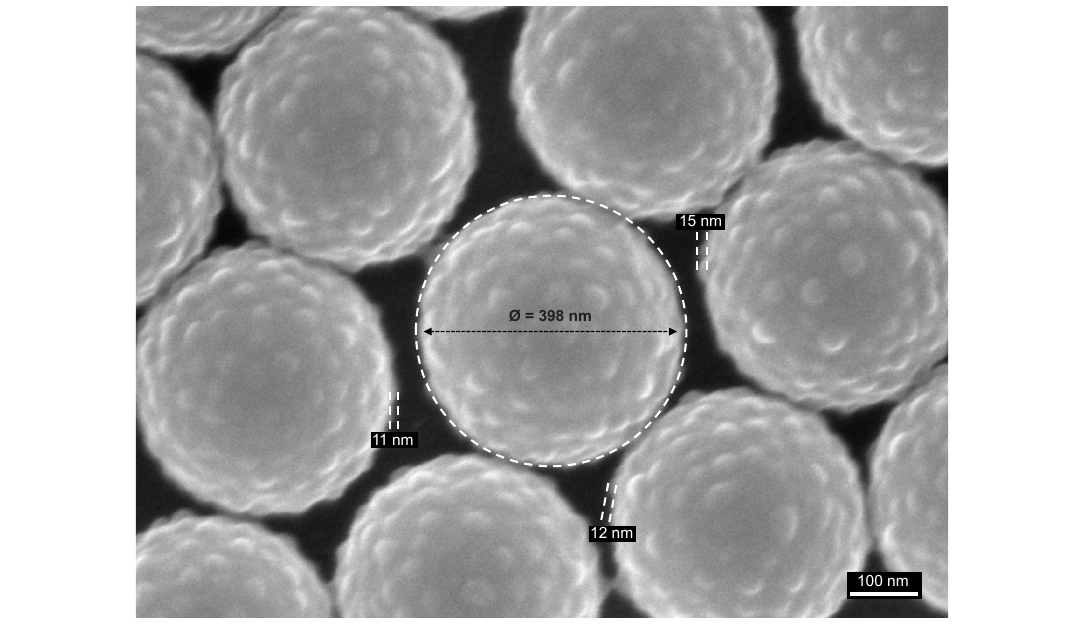} 
\caption{ {\bf Scanning electron microscope image of \SI{400}{\nano\metre} fluorescent nanospheres.} The overall sphere size matches specified value by the producer, but it takes into consideration the thickness of the functional group coating. The functional group layer, which is not fluorescent, corresponds to the pilosities around the nanosphere, which have a size greater than \SI{10}{\nano\metre}. Therefore, the effective fluorescent sphere size is smaller. For the fitting model in fig. \ref{fig2}e, the nanosphere diameters were considered \SI{25}{\nano\metre} smaller (\SI{12.5}{\nano\metre} for each side) to account for the non-fluorescent functional group layer. The image was acquired using a JEOL JSM 6700F field emission scanning electron microscope with a through-lens detector (TLD) and with an electron energy of \SI{3}{\kilo\volt} at a working distance of \SI{6.3}{\milli\metre}.}
\label{SEM_sphere}
\end{figure}

\begin{figure}[h!] 
\centering
\includegraphics[width=1\textwidth]{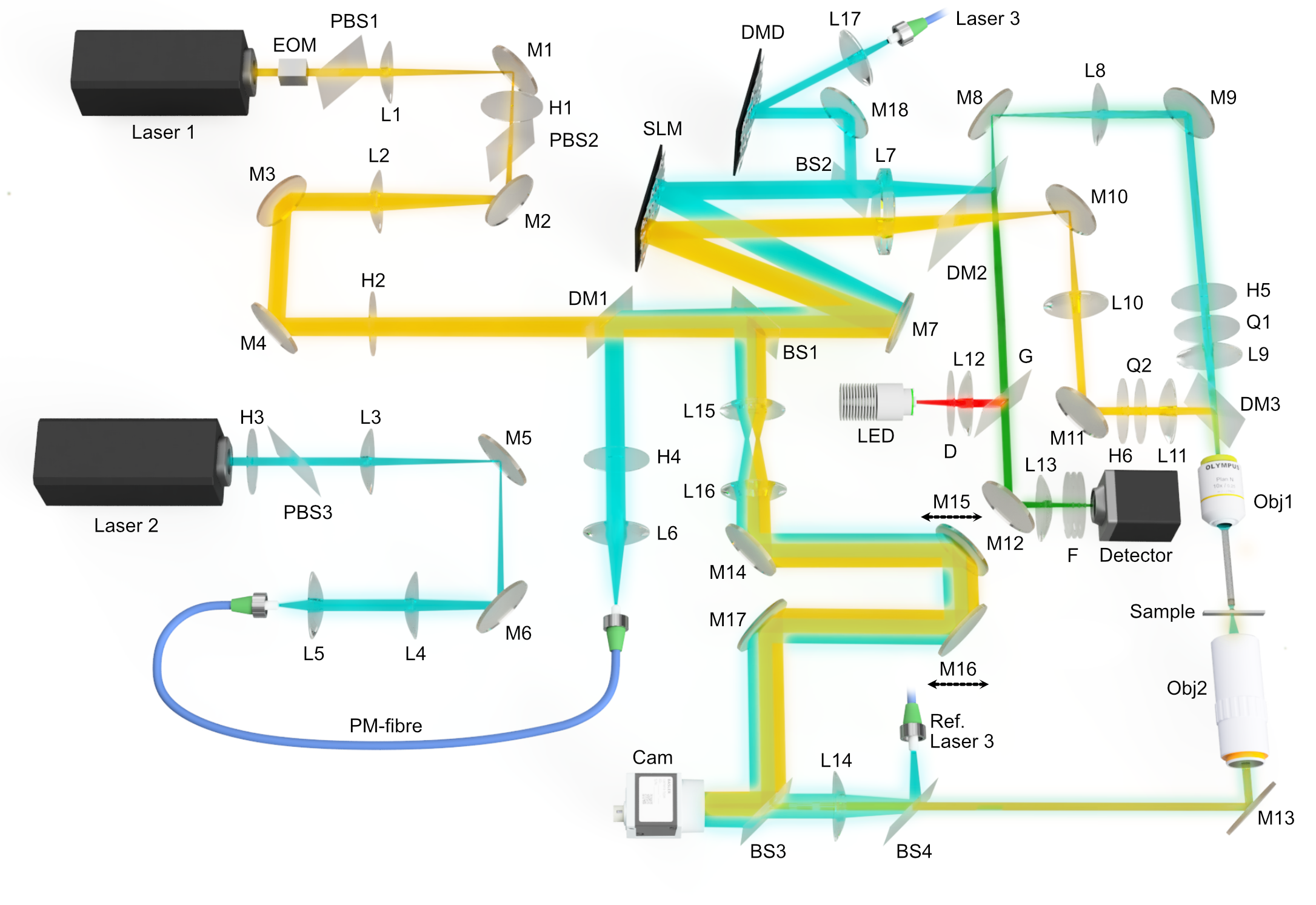} 
\caption{ {\bf The experimental geometry.}
{\bf BS1}: Beamsplitter 10:90 (R:T), Thorlabs BS025.
{\bf BS2}: Beamsplitter 10:90 (R:T), Thorlabs BS025.
{\bf BS3}: Beamsplitter 50:50 (R:T), Thorlabs BS010.
{\bf BS4}: Beamsplitter 50:50 (R:T), Thorlabs BS013.
{\bf CAM}: Camera, Basler ace acA640-750um.
{\bf D}: Ground glass diffuser, Thorlabs DG-10-2200-MD.
{\bf Detector}: Hybrid photomultiplier detector, PicoQuant PMA Hybrid 40.
{\bf DM1}: Dichroic mirror, Thorlabs DMLP505R.
{\bf DM2}: Dichroic mirror, AHF Analysentechnick AG STED Laser Beamsplitter zt 488 RDC.
{\bf DM3}: Dichroic mirror, AHF Analysentechnick AG Shortpass Beamsplitter T 565 SPXR.
{\bf DMD}: Digital micromirror device, ViALUX V-7001.
{\bf EOM}: Resonant electro-optic modulator, QUBIG AM2B-VIS\_0.1.
{\bf F}: Filter set: notch filter, Thorlabs NF488-15; 2x notch filter, Thorlabs NF594-23; Brightline fluorescence filter, Semrock FF01-531/40-25.
{\bf G}: Precision cover glass \#1.5H, Thorlabs CG15KH1.
{\bf H1-H5}: Achromatic half-wave plate, Newport 10RP52-1B.
{\bf L1}: Lens, Thorlabs  AC254-040-A-ML($f=\SI{40}{\milli\metre}$).
{\bf L2}: Lens, Thorlabs  AC254-250-A-ML ($f=\SI{250}{\milli\metre}$).
{\bf L3}: Lens, Thorlabs AC254-035-A-ML ($f=\SI{35}{\milli\metre}$). 
{\bf L4}: Lens, Thorlabs AC254-045-A-ML ($f=\SI{45}{\milli\metre}$).
{\bf L5}: Lens, Thorlabs C240TMD-A ($f=\SI{8}{\milli\metre}$).
{\bf L6}: Lens, Thorlabs AC254-80-A-ML ($f=\SI{80}{\milli\metre}$).
{\bf L7}: Lens, Thorlabs AC254-250-A-ML ($f=\SI{250}{\milli\metre}$).
{\bf L8}: Lens, Thorlabs  AC254-045-A-ML ($f=\SI{45}{\milli\metre}$).
{\bf L9}: Lens, Thorlabs  AC254-150-A-ML ($f=\SI{150}{\milli\metre}$).
{\bf L10}: Lens, Thorlabs  AC254-045-A-ML ($f=\SI{45}{\milli\metre}$).
{\bf L11}: Lens, Thorlabs  AC254-150-A-ML ($f=\SI{150}{\milli\metre}$).
{\bf L12}: Lens, Thorlabs  AC254-040-A-ML ($f=\SI{040}{\milli\metre}$).
{\bf L13}: Lens, Thorlabs  AC254-200-A-ML ($f=\SI{200}{\milli\metre}$).
{\bf L14}: Lens, Thorlabs  AC254-100-A-ML ($f=\SI{100}{\milli\metre}$).
{\bf L15}: Lens, Thorlabs  AC254-050-A-ML ($f=\SI{50}{\milli\metre}$).
{\bf L16}: Lens, Thorlabs  AC254-100-A-ML ($f=\SI{100}{\milli\metre}$).
{\bf L17}: Lens, Thorlabs  AC254-200-A-ML ($f=\SI{200}{\milli\metre}$).
{\bf Laser1}: \SI{592}{\nano\metre} picosecond laser, NKT Katana 06 HP.
{\bf Laser2}: \SI{485}{\nano\metre} picosecond laser, PicoQuant LDH-IB-485-B.
{\bf Laser3}: \SI{488}{\nano\metre} CW laser, Coherent Sapphire SF 488.
{\bf LED}: White LED, Thorlabs MWWHL3.
{\bf M1-M13, M18}: Dielectric mirror, Thorlabs BB1-E02.
{\bf M14-M17}: Optical delay line, Thorlabs ODL100.
{\bf Obj1}: Objective 10x, NA: 0.3, WD: \SI{10}{\milli\metre}, Olympus Plan Fluorite.
{\bf Obj2}: Objective 50x, NA: 0.55, WD: \SI{13}{\milli\metre}, Mitutoyo Plan Apochromat.
{\bf PBS1-PBS3}: Polarising beamsplitter cube, Thorlabs PBS251.
{\bf PM-fibre}: Polarisation-maintaining fibre PM460-HP patch cable, Thorlabs P1-488PM-FC-1.
{\bf Q1-Q2}: Achromatic quarter-wave plate, Newport 10RP54-1B.
{\bf Sample}: Sample placed on a microscope slide holder, Thorlabs MAX3SLH, assembled on a 3-axis motorised stage.
{\bf SLM}: Spatial light modulator, Meadowlark HSP-1920-500-1200.}
\label{setup}
\end{figure}

If a mechanical drift occurs between the excitation and depletion TM acquisition, it causes a displacement between the generated excitation and depletion beams. In such case, a drift correction can be performed by imaging a fluorescent nanosphere with both beams separately, as illustrated in fig. \ref{shift_correction}. The image acquired by scanning the excitation beam is used as a reference (fig. \ref{shift_correction}a). Scanning solely using the depletion beam allowed to visualise the position of the vortex centre, since the image results from the convolution of the fluorescent sphere with the vortex beam (fig. \ref{shift_correction}b). Given the high powers used in the depletion beam, there is still a non-zero probability of exciting fluorescence. The shift between figs. \ref{shift_correction}a and \ref{shift_correction}b was then determined with sub-pixel resolution and applied to the depletion beam, correcting the initial displacement, as shown in fig. \ref{shift_correction}c. 

Note that the drift between the excitation and depletion beam in fig. \ref{shift_correction} was exaggerated on purpose to showcase the correction procedure. Experimentally, the drift observed was typically ranging from sub-pixel to 2 pixels (i.e. smaller than \SI{150}{\nano\metre}).

\clearpage
\section*{Supplementary Figures}

\clearpage
\section*{Suplementary Media}
\renewcommand\thefigure{SM\arabic{figure}}    
\setcounter{figure}{0}    

\setlength{\belowcaptionskip}{-.6cm}
\setlength{\abovecaptionskip}{-.0cm}

\subsection*{Supplementary Video 1 - Impact of time-gate window}
\begin{figure}[h] 
\centering
\includegraphics[width=0.8\textwidth]{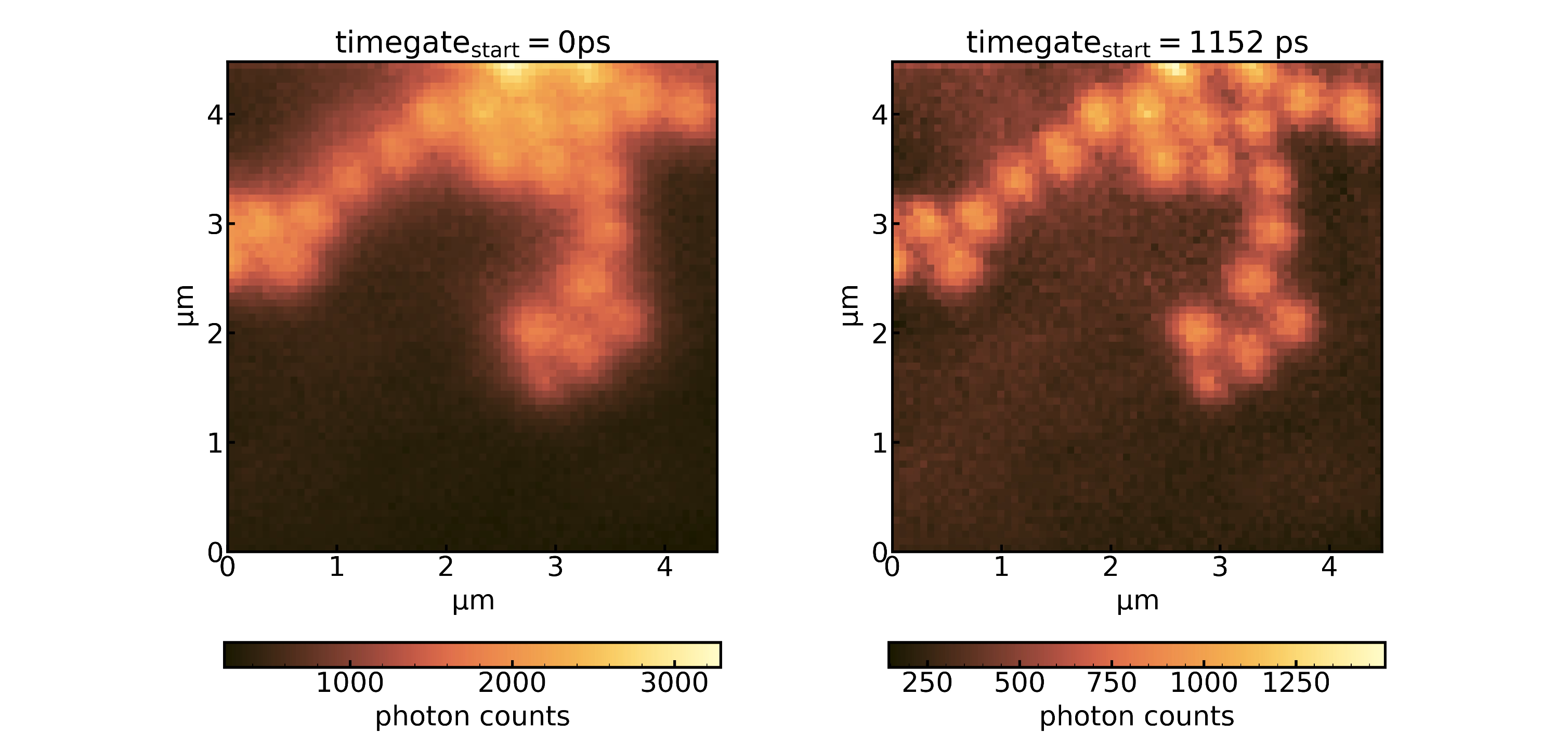} 
\caption{ {\bf Impact of time-gate window starting position.} The initial time-gate window covers the moments before, during, and after depletion. Consequently, it includes excited photons that have not yet been depleted, inducing a reduction in the image contrast. The image contrast enhances once the time-gate window starting position moves forward in time, beyond the depletion event. Following this, only the remaining photons, after depletion has fully occurred, are collected. Naturally, as the temporal starting position moves further away, fewer photons are available for collection, resulting in increased image noise. Note: to better visualise the depletion event in the supplementary video, the depletion pulse was deliberately delayed further from the excitation pulse to increase the temporal separation between them.}
\label{SV1}
\end{figure}

\subsection*{Supplementary Video 2 - Demonstration of imaging procedure}
\begin{figure}[h] 
\centering
\includegraphics[width=0.6\textwidth]{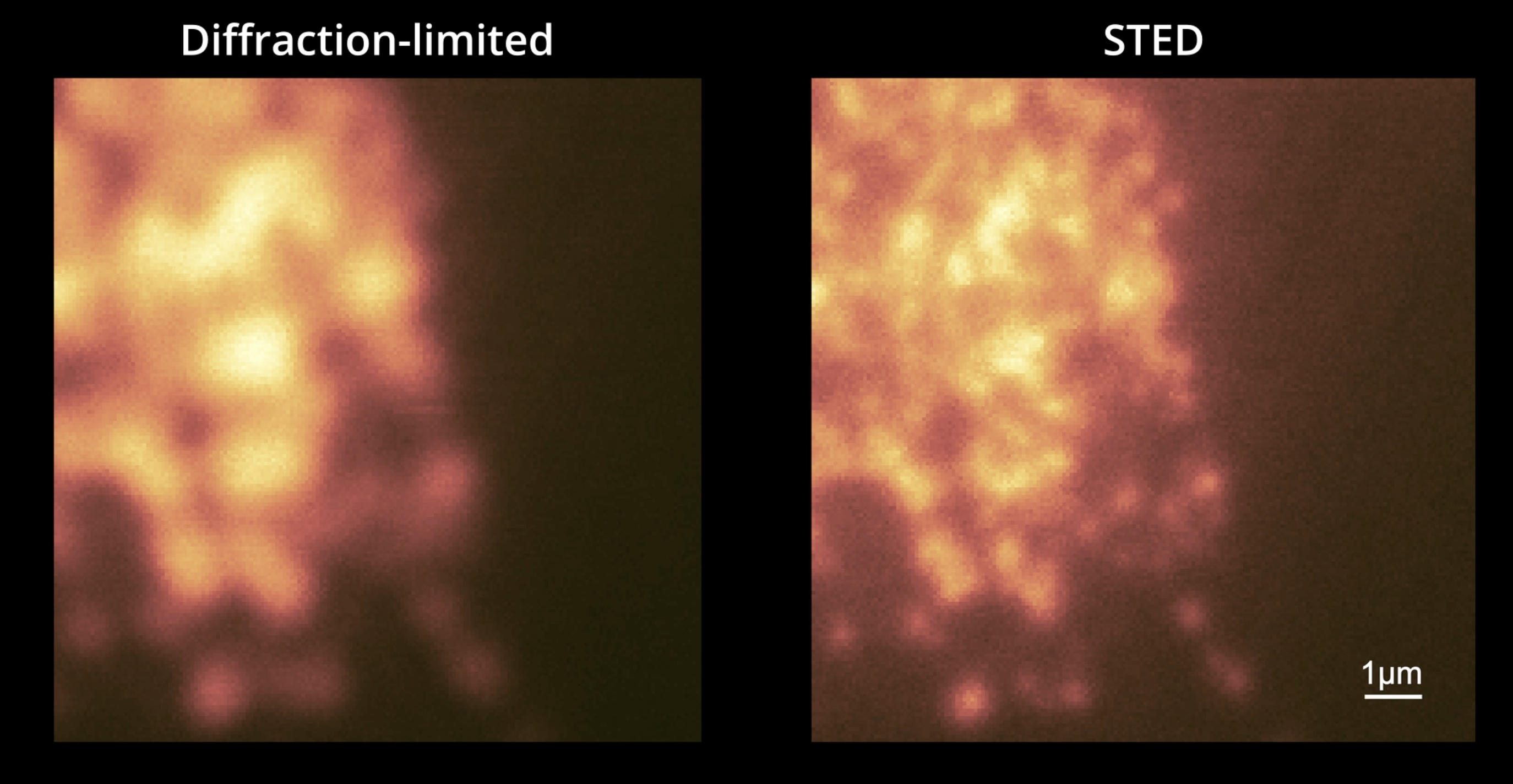} 
\caption{ {\bf Example of imaging procedure.} First, the DMD with a CW laser is used to pre-scan the sample, enabling a fast navigation through the sample, identifying and focussing on a region of interest. This region is then coarsely imaged using the SLM and pulsed excitation in diffraction-limited regime. Afterwards, a region of interest for super-resolution is selected and re-imaged in diffraction-limited regime but with finer sampling. Finally, the same region is re-imaged in STED regime with the same fine sampling. Note that the periodic noise present in the DMD originates from measuring the analog voltage signal obtained directly from the hybrid detector. However, this mode of operation is solely used for rapid navigation and qualitative assessment of the regions of interest.}
\label{SV2}
\end{figure}

\end{document}